\documentclass[preprint,12pt]{elsarticle}

\usepackage{amssymb}
\usepackage{amsmath}
\usepackage{float}
\usepackage{multirow}
\journal{Computational Materials Science}

\begin{document}

\begin{frontmatter}



\title{Multiscale simulations of three-dimensional nanotube networks: Enhanced modeling using unit cells}

\author[AMP]{Fabian Gumpert}
\author[WT,CMP]{Dominik Eitel}
\author[WT,CMP]{Olaf Kottas}
\author[WT,CMP]{Uta Helbig}
\author[AMP]{Jan Lohbreier}

\affiliation[AMP]{organization={Faculty of Applied Mathematics, Physics and Humanities, Technische Hochschule Nürnberg Georg Simon Ohm},Department and Organization
            addressline={Keßlerplatz 12}, 
            city={Nuremberg},
            postcode={90489}, 
            state={Bavaria},
            country={Germany}}

\affiliation[WT]{organization={Faculty of Materials Engineering, Technische Hochschule Nürnberg Georg Simon Ohm},Department and Organization
            addressline={Keßlerplatz 12}, 
            city={Nuremberg},
            postcode={90489}, 
            state={Bavaria},
            country={Germany}}

\affiliation[CMP]{organization={Institute for Chemistry, Materials and Product Development (Ohm-CMP), Technische Hochschule Nürnberg Georg Simon Ohm},Department and Organization
            addressline={Keßlerplatz 12}, 
            city={Nuremberg},
            postcode={90489}, 
            state={Bavaria},
            country={Germany}}
      
%% Abstract
\begin{abstract}
This study presents a simulation approach for three-dimensional nanotube networks using cubic and tetragonal unit cells to enhance modeling efficiency. A random-walk algorithm was developed to generate these networks, which were analyzed in a Finite Element Method (FEM) simulation to assess their electrical conductivity. The percolation probability as a function of the nanotube filling factor can be derived from these simulation results. It is found that smaller tetragonal unit cells can replicate the behavior of larger networks with significantly reduced computational effort, achieving a 20 times reduction in computational time while receiving similar results. In this work, we focus on carbon-doped titanate nanotubes for hydrogen applications, but the method is adaptable for other nanocomposite applications. The findings provide a universal framework for the investigation of nanotube-based materials.
\end{abstract}


%\item Electrical properties of 3D realistic nanotube networks are analyzed by a combination of random-walk algorithm and FEM simulation.
%\item A unit cell approach provides tremendous potential for time reduction.
%\item Size and shape of the unit cell has big influence on the simulation results.

%% Keywords

\end{frontmatter}



\section{Introduction}

Nanostructure network materials are promising candidates for a wide range of next-generation applications. It was found, that nanoparticles and networks formed by them often exhibit new and superior properties in nanocomposite materials. These composite materials consist of nanoparticles, embedded within a matrix. Among the different nanoparticle shapes that can be synthesized, nanotubes are the most interesting for many applications, because of their outstanding properties (e.g. high aspect ratio and high specific surface area) \cite{Zakir2023}. The functional properties of the composite material (e.g. thermal and electrical conductivity or elasticity) are significantly influenced by the network properties, in particular the probability of the network to form contiguous clusters (percolation paths) \cite{Nan2010}. This probability is influenced by morphological parameters, such as the length, diameter, or curvature as well as the relative amount of the nanotubes in the material \cite{Li2007}. These characteristics can be used to tune the functional properties of the nanocomposite according to the requirements of specific applications \cite{Bauer2006, Simi2017, MndezGalvn2021}. Hence, an in-depth understanding of the impact of network properties on the composite material is essential. However, current experimental methods alone are not always able to sufficiently investigate these networks and their influence on the functional behaviour of the material \cite{Zorn2021}. Because of the lack of the experimental methods, theoretical approaches are used to study the impact of tube parameters on material properties. The electrical conductivity of a straight nanotube network are studied with an analytical framework by Ahmadi and Saxena \cite{Ahmadi2024}. Tang et al.\ \cite{Tang2021} proposed an analytical approach to investigate the electrical properties of a network, consisting of sinusoidal-shaped nanotubes. Haghgoo et al.\ combined theoretical considerations and Monte-Carlo simulations to predict the electrical properties of material \cite{Haghgoo2023}. In their work, they investigated the impact of different factors on the electrical conductivity and the piezoresistivity of the nanotube network. However, the validity of theoretical descriptions is still limited, mainly because of geometrical assumptions or physical phenomena which are not included in the theories. Computer aided simulations are conducted to fill the gap and to reveal the influence of the different physical mechanisms on the material behaviour. 

%Nanotube network materials are already used in various applications, such as energy conversion \cite{CARP2004, Ramesh2008, Barnes2010, Bowker2011, Akilavasan2013, Dmitriev2015, Ahmad2015}, nanoelectronics \cite{Chortos2015, Liu2015}, aerospace \cite{Bellucci2007, Gohardani2014}, civil engineering \cite{Siahkouhi2021, Cui2022, Faisal2022, Jannat2023}, medicine \cite{Yao2009, Junkar2020, Zhang2021} and quantum technologies \cite{Baydin2022, Althuon2024}

Classical computer based simulations are deterministic, meaning that for a given set of input variables the output values will be always the same. In contrast, stochastic simulators include random behaviour which can be found in many real-world applications. Zeng et al.\ used Monte-Carlo based stochastic simulations to generate two-dimensional networks of straight nanotubes to study the percolation probability for different tube lengths and alignments \cite{Zeng2011}. Pereira et al.\ \cite{Pereira2011} modeled three-dimensional random nanotube networks to analyze the impact of various network properties on the electrical conductivity of the composite material. For the investigation of the electrical and mechanical properties of wavy nanotubes, three-dimensional simulations are conducted by other groups \cite{Natarajan2019, Stein2015}. Simoneau et al.\ \cite{Simoneau2015} generated three-dimensional models of curved carbon nanotube (CNT) networks with Monte Carlo simulations. They focused their work on the impact of the overlap at junctions and their influence on the electrical conductivity. Kumar et al.\ developed an algorithm to model three-dimensional complex nanotube networks \cite{Kumar2024}.

Due to the multi-scale aspect, the simulation of full three-dimensional nanotube networks is a challenging task and large networks require high-memory computers and result in long computation times. As a consequence, the applicability of these simulations is limited. From other fields, such as solid-state physics, the usage of unit cells is well established in order to reduce the computational effort and consequently, time. Chen et al.\ investigated the stiffness of composite materials for 2D and 3D unit cells \cite{Chen2015}. The authors employ straight and circular nanotubes in their study. Zhang et al.\ \cite{Zhang2020} employed a unit cell modeling approach to investigate the mechanical properties of a 3D periodic array of wavy nanotubes. However, the nanotubes in these works either lack the random nature \cite{Zhang2020} or the realistic curvy geometry \cite{Zhang2020, Chen2015}. Both characteristics are essential for the employed nanotubes in our studies.

The present work proposes the usage of periodic three-dimensional nanotube networks (unit cells) to investigate the functional properties of composite materials. A random-walk algorithm was developed to generate periodic networks, which were analyzed using Finite Element Method (FEM) simulations. From the FEM results, the existence or absence of a percolation path in the network can be concluded. The percolation probability of the network is studied for varying filling factors and different network sizes. We focus on titanate nanotubes for hydrogen applications which determines the properties of nanotubes. Nevertheless, the proposed approach is highly adaptable and can be applied to nanotube networks designed for various applications with differing properties. The developed guidelines are universal and can be utilized across a wide range of applications.
%It was found that smaller unit cells sizes are able to replicate the percolation probabilities of a larger reference domain, reducing computational times by a factor of 20 in our studies.

\section{Material and Methods}

\subsection{Composite material}
%The preparation of titanate nanotubes by hydrothermal treatment is derived from the synthesis of Kasuga et al. \cite{Kasuga1998}. The synthesis was performed in a PTFE flask in an oil bath. In the PTFE flask 2.0 g precursor powder were treated in a 10 M sodiumhydroxide solution. The oil bath was set to a temperature of 130 °C resulting in a temperature of 116 °C inside the flask. The heated alkaline dispersion was constantly stirred at 500 rpm for 24 h.  While the titanate nanotubes are prepared from commercially available TiO\textsubscript{2} (P25, Evonik), the carbon-modified titanate nanotubes  synthesis is based on core-shell C-TiO\textsubscript{2} precursor powder. This precursor is prepared using a one-step gas phase synthesis. The fabrication of the nanotubes is described in detail in Helbig et al. \cite{Helbig2018}. The preparation of the core-shell C-TiO\textsubscript{2} precursor for the carbon-modified titanate nanotubes is described in Eitel et al. \cite{Eitel2023}.
In our investigations, we are interested in a composite material, used as electrodes in proton exchange membrane fuel cells (PEM-FC). Those composite materials are based on a matrix and a filler (catalyst support material). The filler material is synthesized carbon doped titanate nanotubes (C-TNT). Titanate nanotubes were initially discovered by Kasuga et al. in 1998 \cite{Kasuga1998} and attracted a lot of attention due to their unique functional properties (e.g. physio-chemical properties, high surface area and high chemical stability \cite{Varghese2003, Carabin2015, Brunelli2016}). Modifying titanate nanotubes with carbon enables an even broader application spectrum due to high conductivity.

\begin{figure}[H]
\centering
\includegraphics[width=0.45\linewidth]{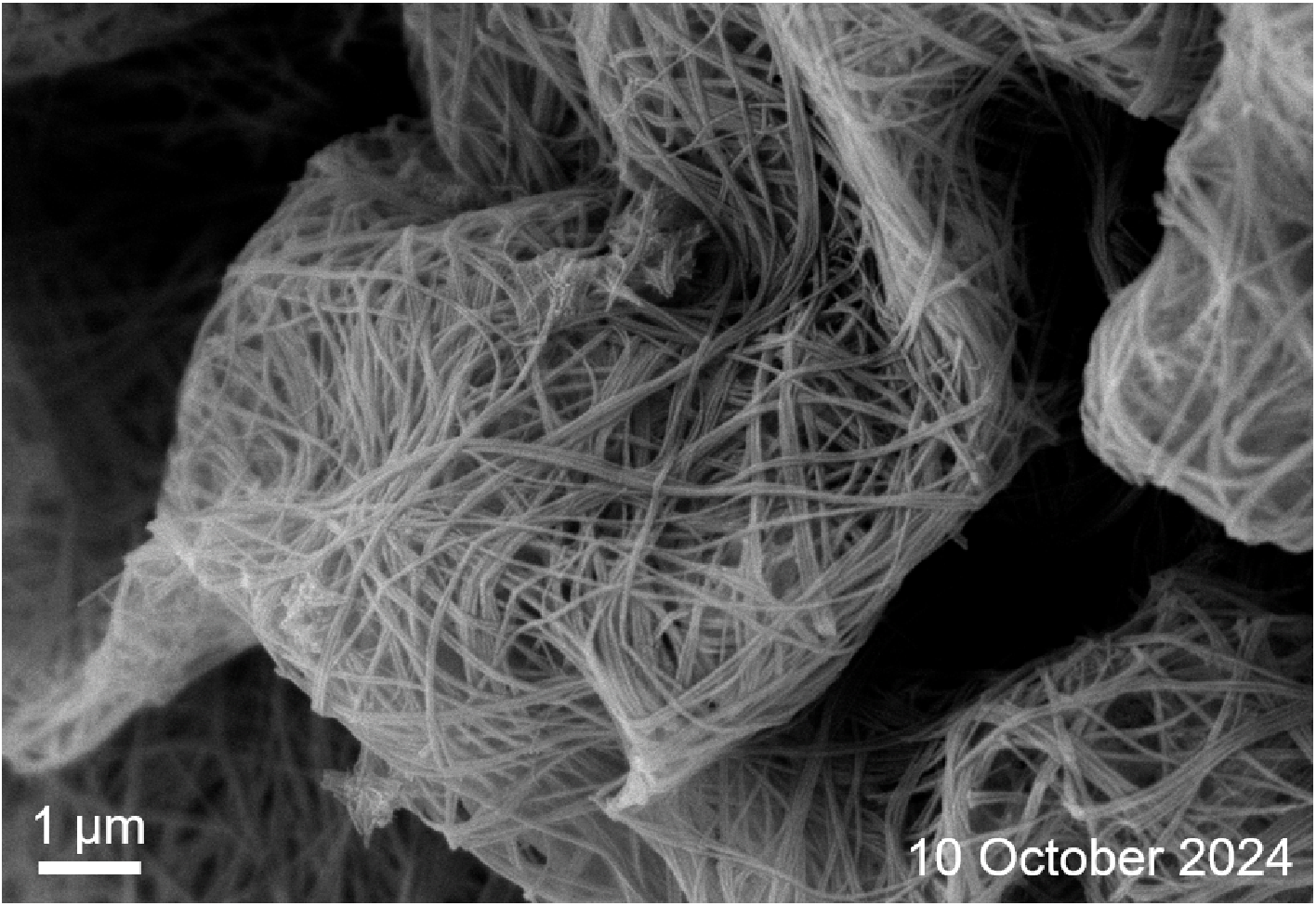}
\caption{SEM image of titandioxide nanotubes after synthesization procedure.}
\label{fig_TNTs}
\end{figure}

The preparation of titanate nanotubes by hydrothermal treatment is derived from the synthesis of Kasuga et al. \cite{Kasuga1998}. The synthesis was performed in a PTFE flask in an oil bath. In the PTFE flask 2.0 g precursor powder were treated in a 10 M sodiumhydroxide solution. The oil bath was set to a temperature of 130 °C, resulting in a temperature of 116 °C inside the flask. The heated alkaline dispersion was constantly stirred at 500 rpm for 24 h.  While the titanate nanotubes are prepared from commercially available TiO\textsubscript{2} (P25,  Acros Organics), the carbon-modified titanate nanotubes  synthesis is based on core-shell C-TiO\textsubscript{2} precursor powder. This precursor is prepared using a one-step gas phase synthesis. The fabrication of the nanotubes is described in detail in Helbig et al. \cite{Helbig2018}. In Eitel et al. \cite{Eitel2023}, the preparation of the core-shell C-TiO\textsubscript{2} precursor for the carbon-modified titanate nanotubes is described. A scanning electron image (SEM) of the resulting carbon-modified titanate nanotubes is shown in Fig. \ref{fig_TNTs}.

In order to obtain conductivity values of the filler material for simulation, 100 mg of nanotubes were pressed in a pellet form with a diameter of 13 mm using uniaxial pressing with 3 kN force. The thickness of the pellets were measured before they were coated on both sides with conductive silver LS200N BC (Hans Wolbring GmbH).

\begin{figure}[H]
\centering
\includegraphics[width=0.45\linewidth]{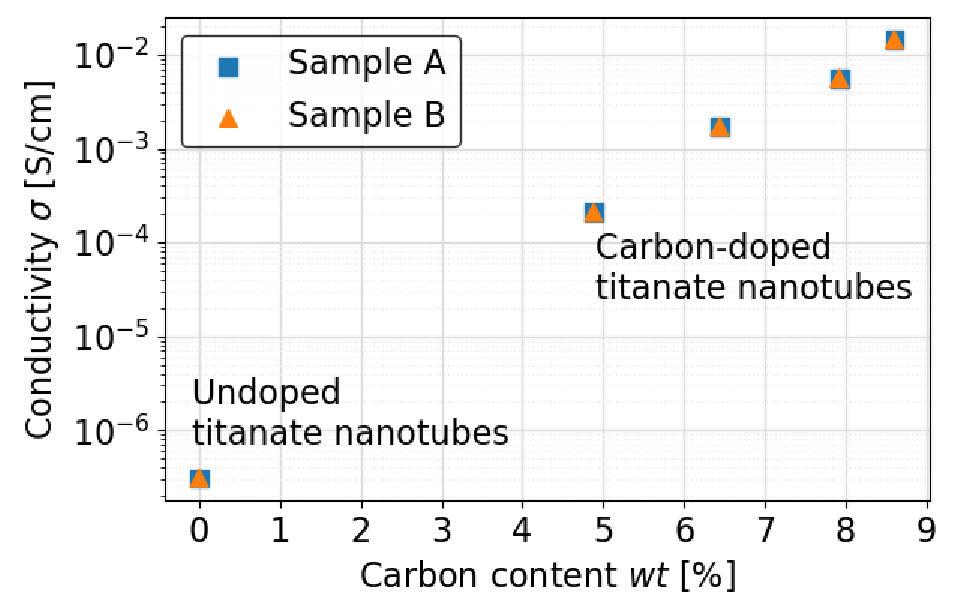}
\caption{Experimentally determined electrical conductivity of carbon doped titanate nanotubes as a function of carbon content.}
\label{fig_condTNT}
\end{figure}

The conductivity of the carbon-doped titanate nanotubes was derived from impedance measurements using a Keysight E4990A impedance Analyzer equipped with the 16047E measuring adapter. A 200 mV sinusoidal voltage with a frequency range from 20 Hz to 30 MHz was applied to the pressed powder pellets. From the Nyquist plots semicircles have been fitted and the diameter of the extrapolated semicircle was used as $Z_{Re}$ \cite{Macdonald2006}. Eq. \ref{eq_conductance_formula} was used to calculate The electrical conductivity $\sigma$ (SI-Unit: S/m) of the samples can be calculated with
%and a 3D-printed sample holder

\begin{equation}
    \sigma = \frac{t}{(Z_{\text{Re}} \cdot A)} ,
\label{eq_conductance_formula}
\end{equation}

\noindent where $t$ is the sample thickness (SI-Unit: m), ${Z_\text{Re}}$ is the real part of the resistance (SI-Unit: 1/S), and $A$ is the cross sectional area of the sample surface (SI-Unit: m\textsuperscript{2}). Fig. \ref{fig_condTNT} shows the conductivity in dependence of the carbon content which was determined by thermogravimetric measurements. 

For comparison, the conductivity of undoped titanate nanotubes were fabricated from unmodified titaniumdioxide powder (P25, Acros Organics). Each nanotube sample was divided and pressed in a separate pellet. The results are labeled as 'Sample A' and 'Sample B' respectively. As the plot shows, both samples revealed consistent conductivity values. For all simulations, the conductivity of the samples $\sigma_{\text{TNT}}$ with the highest carbon (content have been selected (0.0144 S/cm) was used. 

We require the densities of nanotubes and the matrix material for the conversion from volumetric filling factor $v_f$ to weight percentage $wt$ in our investigations. Regarding the density of titanate nanotubes few values are published and those are quite vague. A value between 3.64–3.76 g/cm\textsuperscript{3} was published by Zukalová et al.\ in 2005 \cite{Zukalov2005} and we assumed a value of 3.7 g/cm\textsuperscript{3} in this work. The matrix material is commercially available Nafion, whose density is reported to be 1.95 g/cm\textsuperscript{3} \cite{Oberbroeckling2002}. All relevant material properties are summarized in Tab. \ref{tab_matparas}.

\begin{table}[H]
\centering
\caption{Relevant properties of composite materials used in this study.}
\begin{tabular}{l l l}
Material & Property & Value  \\
\hline
C-TNT   &   Electrical  conductivity   &   0.0144 S/cm \\
C-TNT    &   Density   &   3.7 g/cm\textsuperscript{3} \\
Nafion    &   Density   &   1.95 g/cm\textsuperscript{3}\\
\end{tabular}
\label{tab_matparas}
\end{table}

%The composite material consists of two materials, one for the nanotubes and one for the matrix. Among the available materials for the nanotubes, titandioxide has attracted increasing attention from researchers in recent years due to its unique functional properties (e.g. physico-chemical properties and non-toxicity \cite{Varghese2003, Radecka2008, Carabin2015}). The material was discovered in 1998 by Kasuga et al. \cite{Kasuga1998} and is also known as titanate. In this work, we are interested in the simulation of multi-walled titanate nanotubes (TNTs), synthesized as proposed by Helbig et al. in \cite{Helbig2018}. TNTs have a high aspect ratio, which is beneficial for the conductivity of the network \cite{Pereira2011, Ji2021}. Figure \ref{fig_TNTs} shows TNTs after the synthesis. 

%%For the composite material, the clustered TNTs are separated by an ultrasonic process to obtain individual tubes. In contrast to carbon nanotubes (CNTs), which are often approximated by straight tubes \cite{Ahmadi2024}, individual TNTs form more complex networks due to their flexibility. Although titanate nanotubes have a multi-walled structure in reality, they are approximated by solid tubes in our simulations. The electrical conductivity of TNT is reported to be about 1 S/m. 

%%The density of titanium dioxide is assumed to be 4.23 g/cm³ \cite{FernndezWerner2017}. The density of Nafion 117, which is proposed as the matrix material, is reported to be 1.95 g/cm³ \cite{Oberbroeckling2002}. 

\subsection{Stochastic simulator for generating periodic nanotube networks}

\begin{figure}[H]
\centering
\includegraphics[width=0.9\linewidth]{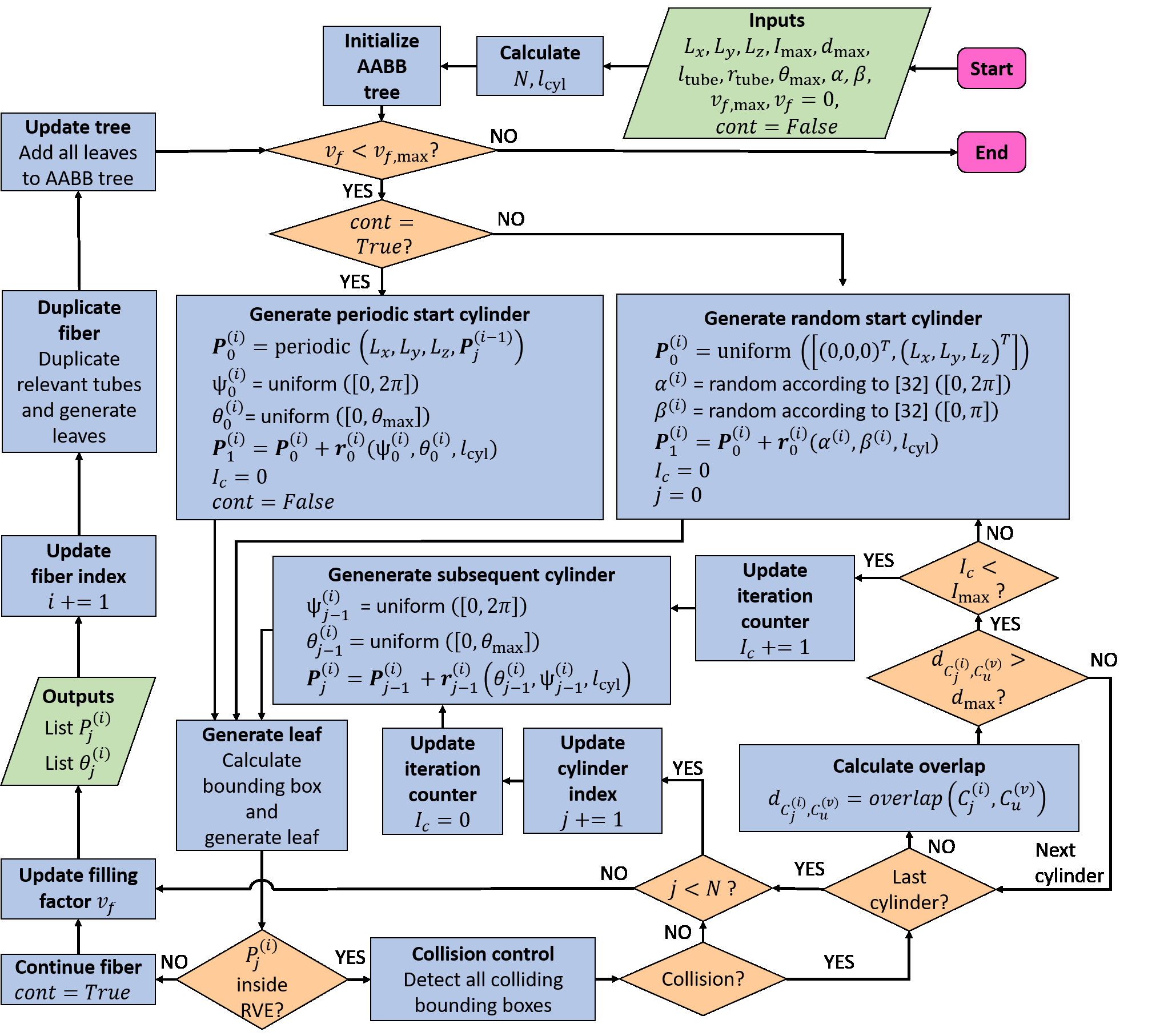}
\caption{Flowchart of the algorithm to generate nanotube networks.}
\label{fig_flowchartcode}
\end{figure}

%The stochastic simulator in this work is based on a random-walk algorithm to generate nanotube networks, similar to the one proposed by Chahal and Adnan \cite{Chahal2020}. However, to make the proposed simulation approach more practical, we improve the computational time for the collision detection. Moreover, we employ a unit cell approach to further reduce the computational time. For this, we modified the proposed algorithm in \cite{Chahal2020} such that we generate periodic nanotube networks. In Fig. \ref{fig_flowchartcode}, the flow chart of the algorithm in this work is illustrated.
The stochastic simulator employs a random-walk algorithm to generate periodic nanotube networks. While the algorithm is based on the work of Chahal and Adnan \cite{Chahal2020}, it was modified to suit the requirements of our work. Necessary changes and additions were made in order to generate periodic nanotube networks which can be used as unit cells in the numerical simulations. An Axis-Aligned Boundary Box (AABB) tree structure reduces the computation time of the algorithm for the collision detection. In Fig. \ref{fig_flowchartcode}, the resulting flow chart of our algorithm presented in this work is illustrated.

% In , Chahal and Adnan propose a random-walk algorithm to generate nanotube networks. We use the algorithm as the basis of our stochastic simulator, but improve the computational time of the collision detection and modify the algorithm such that periodic networks are generated. 

All parameters, used in the algorithm, are summarized and described in Tab. \ref{tab_paras} at end of this section. A representative volume element (RVE) is used as the simulation domain in the later numerical simulations. The RVE, lying within the interval $[\textbf{L}', \textbf{L}' + \Delta \textbf{L}]$, is filled by the algorithm with a periodic nanotube network and is considered a unit cell in this work. In the beginning of the algorithm, different parameters are specified, such as the maximum volumetric filling factor $v_{f,\text{max}}$. In addition, the length of the nanotube $l_{\text{tube}}$ is set to 2000 nm, which is smaller than the typical thickness of an electrode in a PEM-FC, and the radius $r_{\text{tube}}$ of a nanotube, which is set to 12.5 nm. In our investigations, the interval of the RVE is 3D rectangular tetragonal, which is defined by $\textbf{L}' = (0, 0, 0)^T$ and $\Delta \textbf{L} = (L_x, L_y, L_z)^T$. A nanotube is constructed by the successive addition of cylinders. The length of the cylinders ($l_{\text{cyl}} = 2r_{\text{tube}}$) and the number of cylinders $N$ ($N = l_{\text{tube}}/l_{\text{cyl}}$), forming a tube, are derived from the initial parameters. 

Tubes can intersect with previously generated tubes, which are important junctions for the nanotube network. Given that not all tube overlaps are permitted, a collision detection is necessary to control intersections of newly generated cylinders with existing ones. The collision detection in this work consists of two phases. In the first phase, an AABB tree structure \cite{Hart2020} is employed. Introducing an AABB tree structure allows an efficient distance and thus intersection query and is used for a rough, but fast judgment of whether collisions are possible. In the AABB tree structure, each individual cylinder is represented by a rectangular box. The edges of the box are parallel to the axes of the global coordinate system and the box is defined in the tree structure by its limits in each axis ($x_{\text{min}}, x_{\text{max}}, y_{\text{min}} ,y_{\text{max}}, z_{\text{min}}, z_{\text{max}}$). The narrowest possible box enclosing the cylinder is calculated to reduce the number of iterations in the second phase, where the actual distances between two cylinders is calculated to determine whether two cylinders actually intersect. Representing the cylinders in an AABB tree structure in the first phase provides a tremendous advantage in terms of computation time compared to a brute force approach \cite{Meister2021}. 

Before the initialization of a new tube, a while loop compares the current filling factor $v_{f}$ with the maximum filling factor $v_{f,\text{max}}$. The current filling factor is defined as

\begin{equation}
    v_f = \frac{V_{\text{tubes}}}{V_{\text{RVE}}},
\end{equation}

\noindent where $V_{\text{tubes}}$ is the total volume of all nanotubes and $V_{\text{RVE}}$ is the volume of the RVE ($V_{\text{RVE}} = L_x \cdot L_y \cdot L_z$). As long as the current filling factor is below $v_{f,\text{max}}$, additional tubes are created.

The algorithm is implemented in the Python programming language and a cylinder is represented by an object $C_{j}^{(i)}$, where the index $j$ stands for the cylinder number within a tube ($0 \leq j < N $) and the index $i$ indicates the number of the tube. The cylinder object $C_{j}^{(i)}$ is constructed by a line segment, bounded by the points $\bold{P}_{j}^{(i)}$ and $\bold{P}_{j+1}^{(i)}$, respectively. The points are located at the centres of the cylinder bases and a point $\bold{P}_{j}^{(i)}$ is defined by its coordinates $\left(\bold{P}_{j}^{(i)} = \left(x_{j}^{(i)}, y_{j}^{(i)}, z_{j}^{(i)}\right)^T\right)$. Start- and end point are separated by the displacement vector $\bold{r}_{j}^{(i)}$, which can be obtained with

\begin{figure}[H]
\centering
\includegraphics[width=0.45\linewidth]{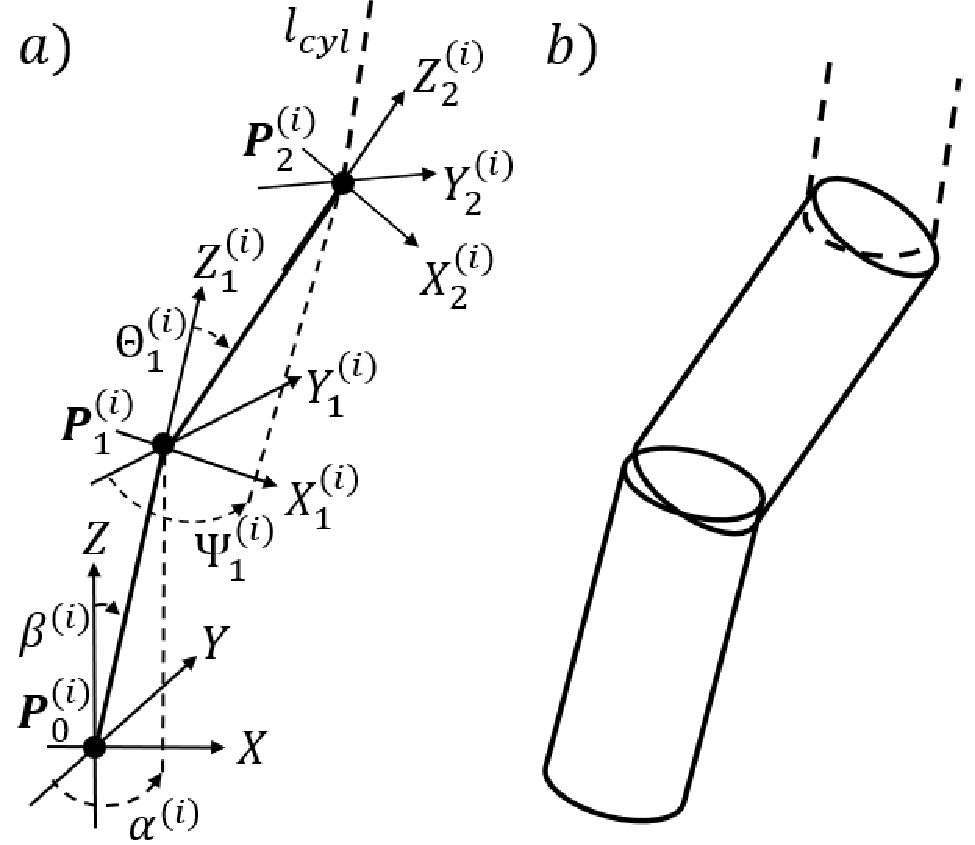}
\caption{a) Illustration of the first two steps to generate a tube with index $i$. From the randomly selected start point $\bold{P}_{0}^{(i)}$, the angles $\alpha^{(i)}$ and $\beta^{(i)}$ define the orientation in the global reference system ($X$, $Y$, and $Z$) of the first cylinder with the end point $\bold{P}_{1}^{(i)}$. The angles $\Psi_{1}^{(i)}$ and $\Theta_{1}^{(i)}$ are selected in the local reference system of point $\bold{P}_{1}^{(i)}$ ($X^{(i)}_1$, $Y^{(i)}_1$, and $Z^{(i)}_1$) and define the orientation of the second cylinder with end point $\bold{P}_{2}^{(i)}$ (adapted from \cite{Chahal2020}). b) First two cylinders of a newly started tube.}
\label{fig_cylinder}
\end{figure}

%%A cylinder is constructed by a line segment which is bounded by the points $\bold{P}_{j,i}$ and $\bold{P}_{j+1,i}$ respectively. These points are located at centres of the cylinder bases. The points are defined by its coordinates (e.g. $\bold{P}_{j,i} = (x_{j,i}, y_{j,i}, z_{j,i})^T$). The points are separated by the displacement vector $\bold{r}_{j+1,i}$, which can be obtained by

\begin{equation}
    \bold{r}_{j}^{(i)} = 
    \begin{pmatrix}
        \cos\left(\Psi_{j}^{(i)}\right) \cdot \sin\left(\Theta_{j}^{(i)}\right)\\
        \sin\left(\Psi_{j}^{(i)}\right) \cdot \sin\left(\Theta_{j}^{(i)}\right)\\
        \cos\left(\Theta_{j}^{(i)}\right) 
    \end{pmatrix}
    \cdot l_{cyl},
    \label{eq_DispVec}
\end{equation}

\noindent where $\Psi_{j}^{(i)}$ and $\Theta_{j}^{(i)}$ are the azimuth and polar angles in the local reference system, respectively (see Fig. \ref{fig_cylinder} a)). From an existing start point $\bold{P}_{j}^{(i)}$, the end point $\bold{P}_{j+1}^{(i)}$ is obtained via

\begin{equation}
    \bold{P}_{j+1}^{(i)} = \bold{P}_{j}^{(i)} + \bold{r}_{j}^{(i)}.
    \label{eq_Displacment}
\end{equation}

\begin{figure}[H]
\centering
\includegraphics[width=0.45\linewidth]{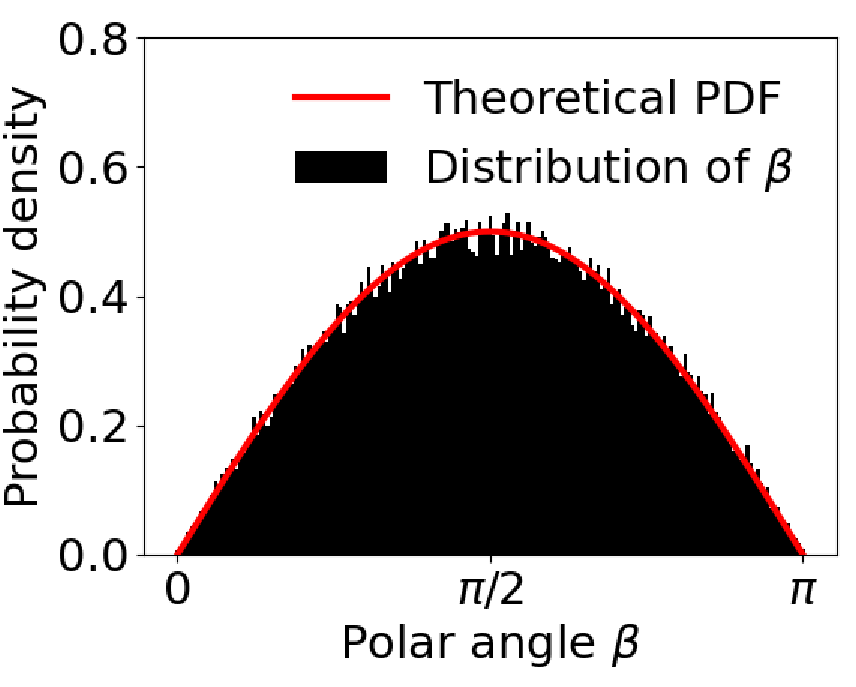}
\caption{The radial angle is randomly selected with the method of Marsaglia (for 100000 samples) and converted to probability densities (black bars). The corresponding theoretical probability density function (PDF) is plotted as solid red line, which is in accordance to Eq. \ref{Eq_Distribution}.}
\label{fig_Dist}
\end{figure}

If the tube with index $i$ is new, the starting point $\bold{P}_{0}^{(i)}$ of the cylinder $C_{0}^{(i)}$ is chosen randomly within the interval of the RVE. For the first cylinder in a new tube, the azimuth angle $\alpha$ and the polar angle $\beta$ define the orientation of the cylinder in the global reference system. In \cite{Chahal2020}, $\alpha$ and $\beta$ are randomly selected uniformly from the interval $[0, 2\pi]$ and $[0, \pi]$, respectively. Uniform randomly selection of $\alpha$ and $\beta$ does, however, result in a higher probability of orientations close to the z-axis, because area elements close to the poles are smaller \cite{Cook1957}. As a consequence, a preferred direction of the tubes is introduced. This is fine for all but the first element of a tube, where no preferred direction is desired. For the generation of the first cylinder, we randomly select the azimuth and polar angles in accordance with \cite{Marsaglia1972}, where Marsaglia proposed a novel method for uniformly sampling on a unit sphere. The probability distribution of the polar angle $P(\beta)$ is 

\begin{equation}
    P(\beta) \propto \text{sin}(\beta) ,
    \label{Eq_Distribution}
\end{equation}

\noindent and is illustrated in Fig. \ref{fig_Dist}.

For the first cylinder of a new tube, the azimuth angle $\alpha$ and the polar angle $\beta$ angle are selected in the global reference frame according to \cite{Marsaglia1972} to achieve a homogeneous orientation for the nanotubes. To compute the displacement vector $\bold{r}_{0}^{(i)}$, $\alpha$ and $\beta$ replace the local polar and azimuth angles in Eq. \ref{eq_DispVec}. The initialization of a random start cylinder finishes with setting both the iteration counter $I_c$ and the cylinder index $j$ to zero. For the generation of subsequent cylinders, $\Psi_{1}^{(i)}$ and $\Theta_{1}^{(i)}$ are selected uniformly. In Fig. \ref{fig_cylinder}, the first two steps of the algorithm are illustrated. The relevant parameters and the resulting cylinders are displayed in Fig. \ref{fig_cylinder} a) and b), respectively.

Even though we assume random orientations of the nanotubes in this work, nanotubes often exhibit a preferred direction in real world applications. The orientation of the tubes can be influenced by either the manufacturing process itself \cite{Beyer2012} or by post-processing methods \cite{Kim2004}. The proposed algorithm is feasible enough to include such effects by modifying the probability distributions for polar and azimuth angle accordingly.

If the tube with index $i$ is the periodic continuation of a previous tube, the coordinates and orientation of the first cylinder $C_{0}^{(i)}$ depend on the last cylinder of the previous tube with index $i-1$. In order to generate periodic structures, the starting point $\bold{P}_{0}^{(i)}$ of the next tube needs to be within the interval $[\textbf{L}', \textbf{L}' + \Delta \textbf{L}]$, again. To achieve this, the starting point can be determined by using the coordinates of the cylinder's end point $\left(\bold{P}_{j}^{(i-1)} = \left(x_{j}^{(i-1)}, y_{j}^{(i-1)}, z_{j}^{(i-1)}\right)^T\right)$. The coordinates of the point $\bold{P}_{0}^{(i)}$ can be calculated with
 
\begin{equation}
    m_{0}^{(i)} = 
    \begin{cases}
        m_{j}^{(i-1)} + L_m & \text{for } m_{j}^{(i-1)} < 0 \\
        m_{j}^{(i-1)} & \text{for } 0 \leq m_{j}^{(i-1)} \leq L_m \\
        m_{j}^{(i-1)} - L_m & \text{for } m_{j}^{(i-1)} > L_m ,
    \end{cases}
\end{equation}

%To guarantee a smooth and continues transition between two unit cells, the local polar angle of the start cylinder is copied from the last cylinder of the previous tube ($\Theta_{j,i} = \Theta_{n,i-1}$). The boolean $cont$ that indicates whether a start cylinder is the continuation of a previous cylinder or not is set to $false$.

\noindent where $m = x, y, z$. For the calculation of the displacement vector (E1. \ref{eq_DispVec}), the azimuth angle $\Psi_{0}^{(i)}$ and the polar angle $\Theta_{0}^{(i)}$ are chosen uniformly random within the local reference frame of point $\bold{P}_{j}^{(i-1)}$. At the end of generating a periodic start cylinder, the iteration counter $I_c$ is set to zero and the boolean $cont$, which indicates if a tube is the periodic continuation of a previous tube, is set to its default value \textit{False}.

Regardless if the tube is new or the periodic continuation of the previous tube, the next step is to create a cylinder object $C_{j}^{(i)}$ and the associated bounding box, which is used to generate a new leaf in the AABB tree structure. A control sequence follows to check whether the new cylinder leaves the RVE or not. 

If the cylinder leaves the RVE ($\bold{P}_{j}^{(i)}$ is outside the RVE), the boolean $cont$ is set to $True$ and the finalization of the tube starts with updating the current filling factor $v_f$. The coordinates of all points in the tube with index $i$ and the corresponding polar angles are exported as text files and the tube index is incremented. Due to the periodicity of the unit cell, relevant cylinders are duplicated around the RVE and their corresponding bounding boxes are calculated for the collision detection. Before the subsequent tube with index $i + 1$ is generated, all duplicated cylinders are added as leaves to the AABB tree. 

If the cylinder is inside the RVE, the collision detection starts. The collision detection with the AABB tree is the first estimation of whether the current tube may intersect with previous tubes by comparing the enclosing bounding boxes. If there are intersecting bounding boxes, the overlap $d_{C_{j}^{(i)},C^{(v)}_{u}}$ between the currently generated cylinder $C_{j}^{(i)}$ and the already existing cylinder $C_{u}^{(v)}$ is calculated for each intersection ($i \neq v$). The next step in the algorithm is the comparison of $d_{C_{j}^{(i)},C^{(v)}_{u}}$ with $d_{\text{max}}$. In this work, a maximum overlap of the cylinders of 20 \% from the tube radius $r_{\text{tube}}$ is allowed, resulting in a value of 2.5 nm for $d_{\text{max}}$. Three different cases for this comparison are possible:

\begin{itemize}
\item[a)] Two bounding boxes intersect, but the tubes do not.
\item[b)] The overlap $d_{C_{j}^{(i)},C^{(v)}_{u}}$ is less than or equal to the maximum allowed overlap $d_{\text{max}}$, which is a permitted intersection between two tubes. These junctions are essential for the macroscopic electrical conductivity of the composite material.
\item[c)] The overlap $d_{C_{j}^{(i)},C^{(v)}_{u}}$ is greater than $d_{\text{max}}$, which is a forbidden intersection.
\end{itemize}

\begin{figure}[H]
\centering
\includegraphics[width=0.45\linewidth]{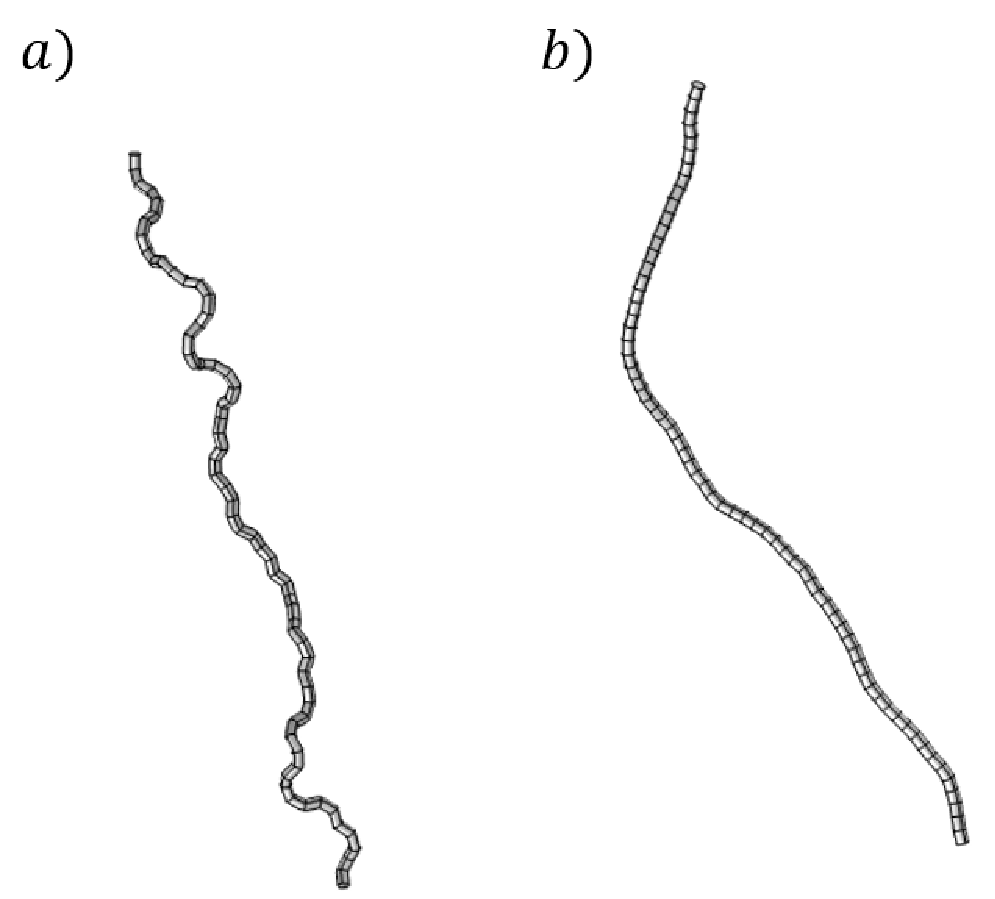}
\caption{a) Tube, generated with local polar angle according to \cite{Marsaglia1972}. b) Tube, generated with uniform randomly selected local polar angle. For both tubes, the interval for the polar angle is $\left[0, \Theta_{\text{max}} \right]$.}
\label{fig_tubetypes}
\end{figure}

%%First, the bounding boxes intersect, but the tubes do not. Second, $d$ is less than or equal to the maximum allowed penetration depth $d_{max}$, which is a permitted junction between two tubes. These junctions are essential for the macroscopic electrical conductivity of the composite material. Both cases are allowed and the algorithm will continue either to check the next bounding box intersection (if there are more collisions) or to continue to generate the next cylinder. Third, $d$ is greater than $d_{max}$ which is a forbidden intersection. 
If a forbidden intersection (case c)) is detected, the iteration counter $I_c$ is compared with the maximum allowed iterations to generate a cylinder $I_{\text{max}}$. As long as $I_c$ is less than $I_{\text{max}}$, the counter is incremented and the current cylinder is replaced by a new cylinder. When the iteration counter reaches $I_{\text{max}}$, the current tube is rejected and a new tube is started. 
The first two cases (a) and b)) are allowed and the algorithm will continue either to check the remaining intersections if there are any further bounding box intersections or to accept the cylinder $C_{j}^{(i)}$ if all intersections are permitted. As long as the cylinder index $j$ is smaller than $N$, a subsequent cylinder is generated. The iteration counter $I_c$ is set to zero and the cylinder index is incremented before the generation. For subsequent cylinders, the local azimuth angle $\Psi_{j}^{(i)}$ and the local polar angle $\Theta_{j}^{(i)}$ are chosen uniformly random from the interval $\left[0, \Theta_{\text{max}} \right]$ and $[0, 2\pi]$, respectively. In contrast, selecting the local radial angle $\Theta_{j}^{(i)}$ in accordance with \cite{Marsaglia1972} results in an unrealistically wavy nanotube, which is shown in Fig. \ref{fig_tubetypes}.

%If all intersections are permitted, the cylinder $C_{j}^{(i)}$ is accepted. When the cylinder index $j$ reached its maximum value ($N-1$), the finalization of the tube starts. If a forbidden intersection is found, the cylinder index $j$ is incremented and the iteration counter $I_c$ is set to zero, followed by the generation of a subsequent cylinder.

When the cylinder index $j$ reaches its maximum value ($N-1$), the finalization of the tube starts. Tubes are generated until the maximal filling factor $v_{f,\text{max}}$ is reached and the algorithm terminates.

In Tab. \ref{tab_paras}, all parameters of the algorithm are summarized.

\begin{table}[H]
\centering
\caption{Parameters used in the algorithm.}
\begin{tabular}{l l l}
Parameter & SI-Unit & Description  \\
\hline
$L_x$    &   m   &   Length of the RVE in x-direction.\\
$L_y$    &   m   &   Length of the RVE in y-direction.\\
$L_z$    &   m   &   Length of the RVE in z-direction.\\
$I_{\text{max}}$ & 1 & Maximum iterations to find a subsequent cylinder.\\
$l_{\text{tube}}$    &   m   &  Length of the nanotubes.\\
$r_{\text{tube}}$    &   m   &   Radius of the nanotubes.\\
$\Theta_{\text{max}}$ & rad & \begin{tabular}{@{}l@{}}Maximal polar angle for the generation of a \\subsequent cylinder.\end{tabular}\\
$\alpha$ & rad & Polar angle in the global reference system. \\
$\beta$ & rad & Azimuth angle in the global reference system. \\
$d_{\text{max}}$    &   m   &   Maximum overlap between two cylinders. \\
$d_{C_{j}^{(i)},C^{(v)}_{u}}$  & m & \begin{tabular}{@{}l@{}} Overlap between cylinder $j$ in tube $i$ \\ and cylinder $u$ in tube $v$.\end{tabular}  \\
$v_{f,\text{max}}$ & 1 & \begin{tabular}{@{}l@{}}Maximum ratio of nanotube volume to overall\\ RVE volume.\end{tabular}\\
$v_f$ & 1 & Filling factor of nanotubes to overall RVE volume.\\
$N$ & 1 & Number of cylinders in a tube.\\
$l_{\text{cyl}}$ & m & Length of one cylinder. \\
$i$ & 1 & Tube index.\\
$j$ & 1 & Cylinder index.\\
$I_{c}$ & 1 & Iteration counter for finding a subsequent cylinder.\\
$\Theta^{(i)}_{j}$ & rad & Polar angle in the local reference system.\\
$\Psi^{(i)}_{j}$ & rad & Azimuth angle in the local reference system.\\
$\textbf{P}^{(i)}_{j}$ & [m,m,m]$^T$ & Coordinate of start point for cylinder $j$ in tube $i$.\\
$\textbf{r}^{(i)}_{j}$ & [m,m,m]$^T$ & Displacement vector from point $\textbf{P}^{(i)}_{j}$ to $\textbf{P}^{(i)}_{j+1}$.\\

\end{tabular}
\label{tab_paras}%
\end{table}

\subsection{Generation of unit cells}
\begin{figure}[H]
\centering
\includegraphics[width=0.7\linewidth]{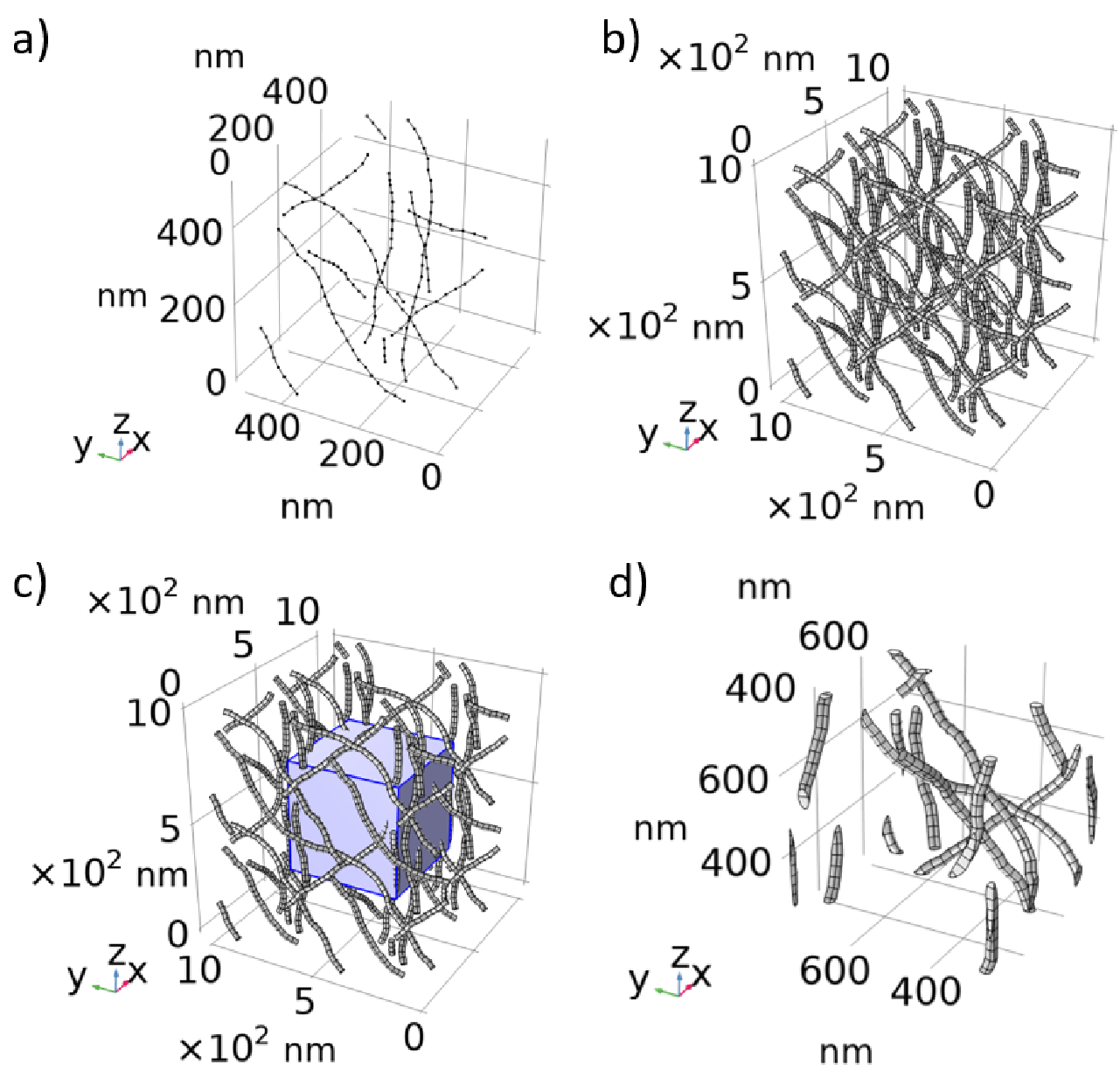}
\caption{Steps to generate a unit cell from periodic nanotube network, illustrated for a cubic unit cell with edge length of 500 nm. a) Import of the periodic nanotube network, generated from our algorithm. b) Duplication of the network and creating 3D tubes. c) Extraction of the central nanotubes by an Intersection operation with the nanotubes and the blue highlighted cube. d) Finalized simulation domain for the numerical evaluation.}
\label{fig_genUnitCell}
\end{figure}

The nanotube networks, generated by the random-walk algorithm, are used to build unit cells. The coordinates of the tubes are imported in COMSOL Multiphysics\textsuperscript{\tiny\textregistered} as vertices of a polygon (see Fig. \ref{fig_genUnitCell} a)). Every polygon is replicated in a three-dimensional array (initial polygon is copied and shifted by $(L_x, 0, 0)^T$, $(0, L_y, 0)^T$, $(0, 0, L_z)^T$, $(L_x, L_y, 0)^T$, $(L_x, 0, L_z)^T$, $(0, L_y, L_z)^T$, and $(L_x, L_y, L_z)^T$). The polygons of tubes, which are the continuation of a previous tube, share a start- and endpoint, respectively. These polygons are concatenated to avoid inconsistencies in the simulation domain. The next step is the generation of three-dimensional objects for the later numerical simulation. For every polygon, a work plane is defined which is perpendicular to the first segment of the polygon. An arbitrary 2D crossectional shape, centered at the vertex, is generated in  the work plane. In our case, this shape is a circle with radius $r_{\text{tube}}$. For all polygons, a \textit{Sweep} operation is performed along the polygon to generate three-dimensional objects (see Fig. \ref{fig_genUnitCell} b)). Furthermore, a tetragonal in the interval $\left[ \left(\frac{1}{2} L_x, \frac{1}{2} L_y, \frac{1}{2} L_Z)^T, (\frac{3}{2} L_x, \frac{3}{2} L_y, \frac{3}{2} L_Z \right)^T \right]$ is used to extract the tubes in this volume by an \textit{Intersection} operation (see Fig. \ref{fig_genUnitCell} c)). The simulation domain for the numerical modeling is finalized by applying the \textit{Form Union} operation in COMSOL (see Fig. \ref{fig_genUnitCell} d)) to create a union from all nanotube objects.

\subsection{Numerical modeling}
\begin{figure}[H]
\centering
\includegraphics[width=0.5\linewidth]{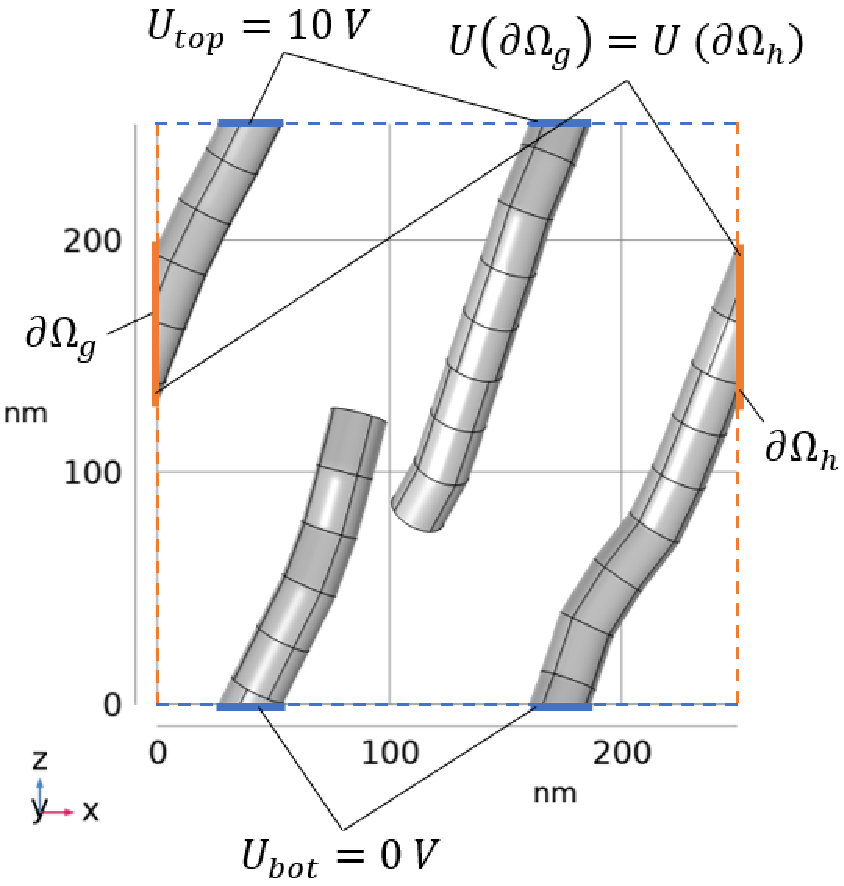}
\caption{Illustration of the relevant boundary conditions at the outer interfaces of the RVE projected to two dimensions. On top and bottom interface, the voltages $U_{top}$ and $U_{bot}$ are defined.}
\label{fig_per_bound_cond}
\end{figure}

The COMSOL Multiphysics\textsuperscript{\tiny\textregistered} platform is employed to determine the electrical conductivity of the nanotube networks. In the electrostatic modeling, the matrix is assumed to be a perfect insulator. As a perfect insulator does not affect the electrical conductivity of the composite material for DC currents, the matrix domain is excluded from the numerical simulations. As a consequence, the existence of a percolation path from top to bottom electrode can be determined by comparing the simulated conductivity of the unit cell with a suitable threshold. For the determination of the electrical conductivity, the electrostatic case is assumed, where Maxwell's equations can be simplified. In the simulation domain $\Omega$, a local formulation of Ohm's Law

\begin{equation}
    \bold{J} = \sigma \cdot \bold{E},
\end{equation}

\noindent with $\bold{J}$ being the current density (SI-unit: A/m\textsuperscript{2}), $\sigma$ being the electrical conductivity (SI-Unit: S/m), and $\bold{E}$ being the electric field (SI-unit: V/m) is solved for. We assume isotropic material properties $\sigma$ and thus, reduces to a scalar. In the electrostatic case, the electric field $\bold{E}$ can be calculated as

\begin{equation}
    \bold{E} = - \nabla U,
\end{equation}

\noindent where $U$ is the electric potential (SI-unit: V). For our analysis, a potential of 10 V at all tube boundaries which intersect with the upper interface of the RVE (top electrode) is applied. A ground potential is applied to all boundaries which intersect with the lower interface of the RVE (bottom electrode). To model the periodic behaviour of the unit cell, a continuity boundary condition is applied to every surface which intersects with the limits of the RVE in x- or in y-direction. In Fig. \ref{fig_per_bound_cond}, the geometry of a unit cell nanotube is shown. 

Boundaries are highlighted where the continuity boundary condition in x-direction is applied. As a consequence of the periodic nature of the unit cell, tubes which leave the RVE will enter the RVE from the other side. Thus, the boundaries, intersecting with the borders in xz- and yz- planes of the RVE, are generated pairwise $\partial \Omega^{(1)}_k$ and $\partial \Omega^{(2)}_k$, respectively. The continuity boundary condition is defined as

\begin{equation}
    U(\partial \Omega^{(1)}_k) = U(\partial \Omega^{(2)}_k) ,
\end{equation}

%\noindent where $U$ is the electric potential at boundary $\partial \Omega_g$ and $U_h$ is the electric potential at boundary $\partial \Omega_h$. 

\noindent with $k$ being the boundary pair index. An electric insulation boundary condition is applied to the remaining boundaries of the nanotubes networks

\begin{equation}
    \bold{n} \cdot \bold{J} = 0,
\end{equation}

\noindent where $\bold{n}$ is a normal vector on the boundary $\partial \Omega$.

%All calculations are computed on an Intel(R) Xeon(R) Gold 6154 CPU workstation with 512 GB RAM.
An AMD Ryzen Threadripper PRO 5965WX 24-Cores and an AMD Ryzen Threadripper PRO 7975WX 32-Cores were used to compute calculations. While differences in computation times between these two machines are negligibly small, the large working memory (512 RAM) of the second work station was necessary to perform the computations with the reference unit cell and with high filling factors.

\subsection{Weibull distribution}
It was found that the percolation probability as the function of the filling factor $v_f$ matches the cumulative probability function (CDF) of a Weibull distribution \cite{Pfeifer2010, Plyushch2020}. The two-parameter Weibull distribution is a continuous probability distribution, which is characterized by the shape parameter $\lambda$ and scaling parameter $k$ \cite{Weibull1951ASD}. %The Weibull probability density function (PDF) is defined as

%\begin{equation}
%    P(v_f; \lambda, k) = \frac{k}{\lambda} \left( \frac{v_f}{\lambda} \right)^{(k-1)}\cdot e^{-(v_f/\lambda)^k} ,
%\end{equation}

%\noindent for $v_f \geq 0$. The Weibull distribution is used in many engineering applications, ranging from material science, electronics to aerospace \cite{Luko1999}. 

The Weibull CDF is used to characterize the percolation probability and can be written as

\begin{equation}
    CDF_{W}(v_f; \lambda, k) = 1-exp[-(v_f/\lambda)^k] .
    \label{eq_CDF}
\end{equation}

The parameters $\lambda$ and $k$ are fitted to the data sets by the least squares method and are used to compare the percolation probability of nanotube networks with different unit cells.

\section{Results and Discussion}
For nanotube composite materials, the probability of the network to form a percolation path is essential for many macroscopic properties of the material. Due to its importance, this probability is used to evaluate whether a small unit cell can represent a targeted RVE or not. In the following investigations, the targeted RVE is a cube with an edge length of 2500 nm. This dimension is chosen because the thickness of the electrode layer for PEMFCs is normally around 2500 nm. The existence of a percolation path can be determined by comparing the simulated conductivity $\sigma$ with a threshold $\sigma_{\text{th}}$. Since no matrix material is included in the simulations, the conductivity of the unit cell equals 0 S/cm if no percolation path exists. However, the conductivity is typically below $10^{-14}$ S/cm in this case due to numerical errors. When the nanotubes form a network from top to bottom of the RVE, the simulated conductivity increases by multiple orders of magnitude. From the results, a threshold $\sigma_{\text{th}}$ of $10^{-11}$ S/cm has proven suitable to determine whether a nanotube network forms a percolation path or not. In addition, a theoretical upper limit for the conductivity $\sigma_{\text{lim}}$ as a function of the filling factor $v_f$ can be defined as

\begin{equation}
    \sigma_{\text{lim}} (v_f) = \sigma_{\text{TNT}} \cdot v_f. 
\end{equation}

In theoretical publications, the filling factor $v_f$ is used to express the ratio of filler to composite material \cite{Tang2021, Haghgoo2023, Chahal2020}. Whereas, in experimental work, the components of the material are quantified in accordance to their respective weights. Thus here, the weight percentage $wt$ is employed. to quantify the nanotube content of the composite material, as in the work of Zeng et al. \cite{Zeng2011}. With our work, we want to address both theoretical and experimental groups. Accordingly, our analyses use the filling factor $v_f$ and the weight percentage $wt$. The weight percentage can be derived from the filling factor

\begin{equation}
    wt(v_f) = \frac{v_f \cdot \rho_{\text{tubes}}}{(1-v_f) \cdot \rho_{\text{matrix}}} \cdot 100\% ,
\end{equation}

\noindent where $\rho_{\text{tubes}}$ is the density of the tube material and $\rho_{\text{matrix}}$ is the density of the matrix material (see Tab. \ref{tab_matparas}). 

\subsection{Cubic unit cells}
The results of the electrostatic simulations are presented to determine the influence of the unit cell size on the percolation probability. In this section, the electrical conductivity is determined by numerical simulations for three cubic unit cells, for a filling factor $v_f$ ranging from 0.0 \% (0.0 $wt$ $\%$) to 2.0 \% (4.39 $wt$ $\%$). The dimensions of the unit cells are 2500 nm x 2500 nm x 2500 nm, 1250 nm x 1250 nm x 1250 nm, and 500 nm x 500 nm x 500 nm, respectively. Simulation results, obtained with the largest unit cell are considered to be the most reliable and are taken as reference. In order to enhance the readability of the presented results, abbreviations for the different unit cell sizes are introduced. The unit cell with the dimensions 2500 nm x 2500 nm x 2500 nm is designated as $\text{UC}_{\text{ref}}$. The abbreviations $\text{UC}^{(1)}_{\text{cubic}}$ and $\text{UC}^{(2)}_{\text{cubic}}$ are used for the unit cells with dimensions 1250 nm x 1250 nm x 1250 nm and 500 nm x 500 nm x 500 nm, respectively. The symmetry of the cubic unit cells is used such that 90°-rotations around the x- or y-axis, respectively, are also used for the numerical studies.

%In all simulation results, the existence of a percolation path in the regarding network is derived and the percolation probability as a function of the filling factor $v_f$ is determined. The parameters of the Weibull CDF are used to compare the results of the different unit cell sizes. 

\begin{figure}[H]
\centering
\includegraphics[width=\linewidth]{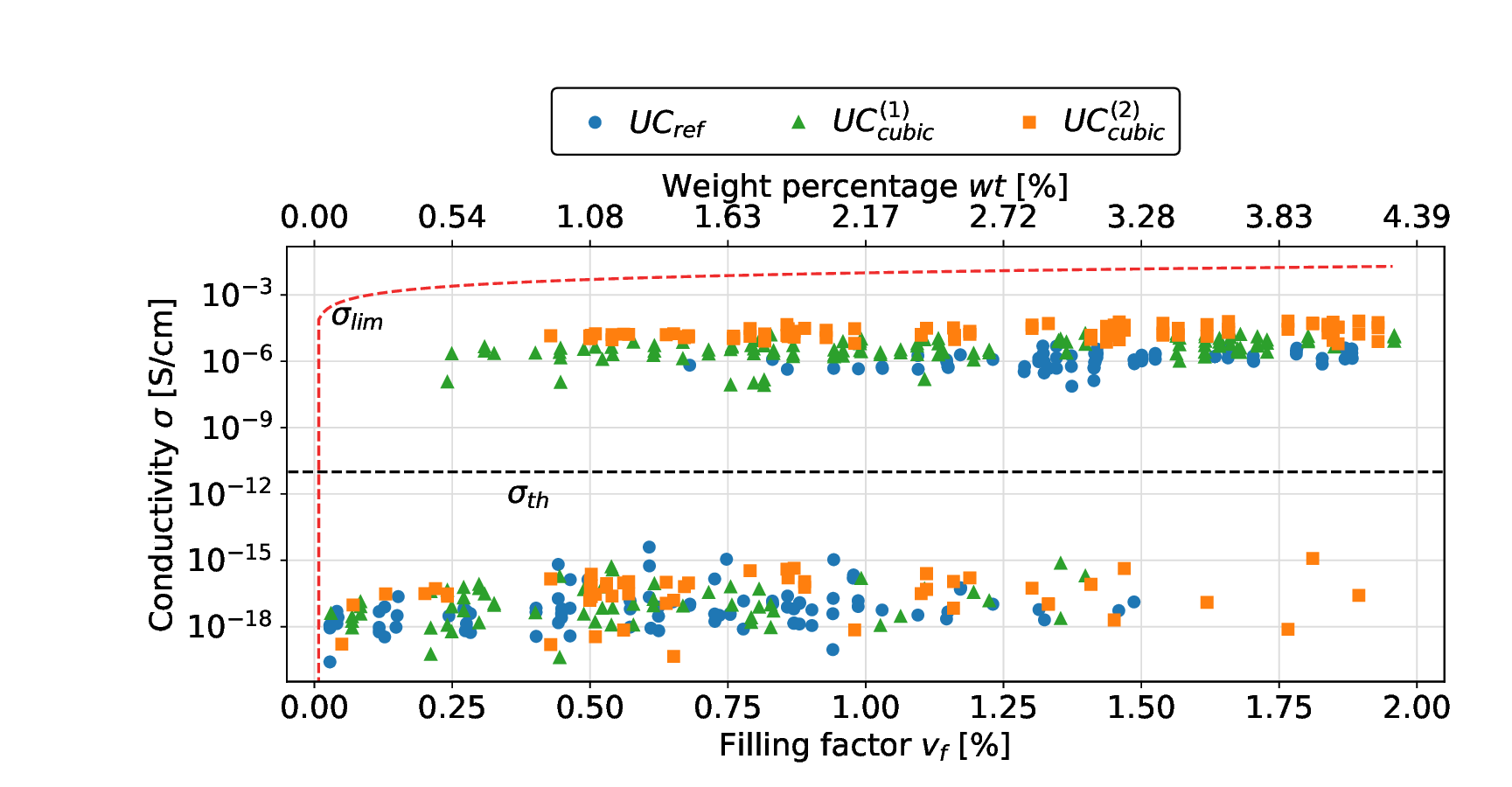}
\caption{Simulated conductivity values for the investigated cubic unit cell sizes as a function of the filling factor $v_f$ (lower x-axis) and the weight percentage $wt$ (upper x-axis), respectively. The simulation results for the unit cells $\text{UC}_{\text{ref}}$ (blue circles), $\text{UC}^{(1)}_{\text{cubic}}$ (green triangles), and $\text{UC}^{(2)}_{\text{cubic}}$ (orange squares) are shown. A red dashed line represents the upper theoretical limit $\sigma_{\text{lim}}$ to the conductivity. The threshold $\sigma_{\text{th}}$, which is used to distinguish whether a percolation path in the network exists or not, is shown as a black dashed line.}
\label{fig_condscubic}
\end{figure}

In Fig. \ref{fig_condscubic}, the simulated conductivities for three different unit cell sizes are presented. For every unit cell size, the data set consists of 150 randomly chosen data points. The electrostatic simulations with the unit cell $\text{UC}_{\text{ref}}$ show, that for filling factors $<$ 0.70 \% (1.52 $wt$ $\%$), the nanotubes do not form a connection between the top and the bottom electrode. An individual nanotube is shorter than the height of the unit cell. As a consequence, it is necessary that at least two tubes have to be connected at a junction to bridge the distance from top to bottom electrode for an conductive path. For filling factors $>$ 1.10 \% (2.39 $wt$ \%), the nanotubes have a high probability to form a percolation path from the top to the bottom. The selection of $\alpha$ according to \cite{Marsaglia1972} results in a isotropic orientation of the tubes and thus to a homogeneous conductivity in x-, y-, and z-direction. Between 0.70 \% (1.52 $wt$ $\%$) and 1.10 \% (2.39 $wt$ $\%$), a "transition phase" exists, where no clear tendency can be observed and which marks the change in probability from obtaining a non-conductive to a conductive network.
%Due to the smaller height of the unit cells, connections from top to bottom electrode occur, even for relatively low filling factors

The results from the FEM simulations with $\text{UC}^{(1)}_{\text{cubic}}$ and $\text{UC}^{(2)}_{\text{cubic}}$ show a similar behaviour, which differs from the results of the reference unit cell. The transition phases for $\text{UC}^{(1)}_{\text{cubic}}$ and $\text{UC}^{(2)}_{\text{cubic}}$ begins already at 0.25 \% (0.54 $wt$ \%) and 0.40 \% (0.87 $wt$ \%), respectively. For both unit cell sizes, the transition phase goes up to 1.40 \% (3.04 $wt$ \%), indicating a relative anisotropical conductivity. Two effects are identified, leading to the observed behaviour. On the one hand, a single nanotube is sufficient to connect the top and bottom electrode without the necessity of an inter-tube junction. On the other hand, the uniform selection of the local radial angle $\Theta$ results in relatively straight tubes (see Fig. \ref{fig_tubetypes} b)). Both effects favour the existence of a percolation path. However, the networks do not reliably form a connection for the investigated filling factors, which is a result of the second phenomenon. In $\text{UC}^{(2)}_{\text{cubic}}$, nanotubes have a relatively large volume in contrast to nanotubes in $\text{UC}_{\text{ref}}$. As a consequence, few nanotubes in $\text{UC}^{(2)}_{\text{cubic}}$ result in an equivalent filling factor as many nanotubes in $\text{UC}_{\text{ref}}$. Hence, few relatively straight nanotubes in $\text{UC}^{(2)}_{\text{cubic}}$ result in a more anisotropic electrical conductivity of the unit cells.

In order to characterize the percolation probability, the results, presented in Fig. \ref{fig_condscubic} are used to conclude whether the utilized networks exhibit a percolation path or not. The conductivity of networks which connect top and bottom electrode is $>$ 10\textsuperscript{-8} S/cm in all simulations. In contrast, networks, which do not exhibit such a connection, show conductivities below 10\textsuperscript{-14} S/cm. The threshold value is chosen to lie at the midpoint of these values ($\sigma_\text{th}$ equal to 10\textsuperscript{-11} S/cm). Applying the threshold to the data in Fig. \ref{fig_condscubic} allows to differentiate between networks that exhibit a percolation path and those that do not.

%In $\text{UC}^{2}_{\text{cube}}$, a single tube segment has the filling factor $9.82\cdot 10^{-5}$ \%, whereas $v_f$ for a segment in the unit cell $\text{UC}_{\text{ref}}$ is equal to $7.85\cdot 10^{-7}$ \%. 

\begin{figure}[H]
\centering
\includegraphics[width=\linewidth]{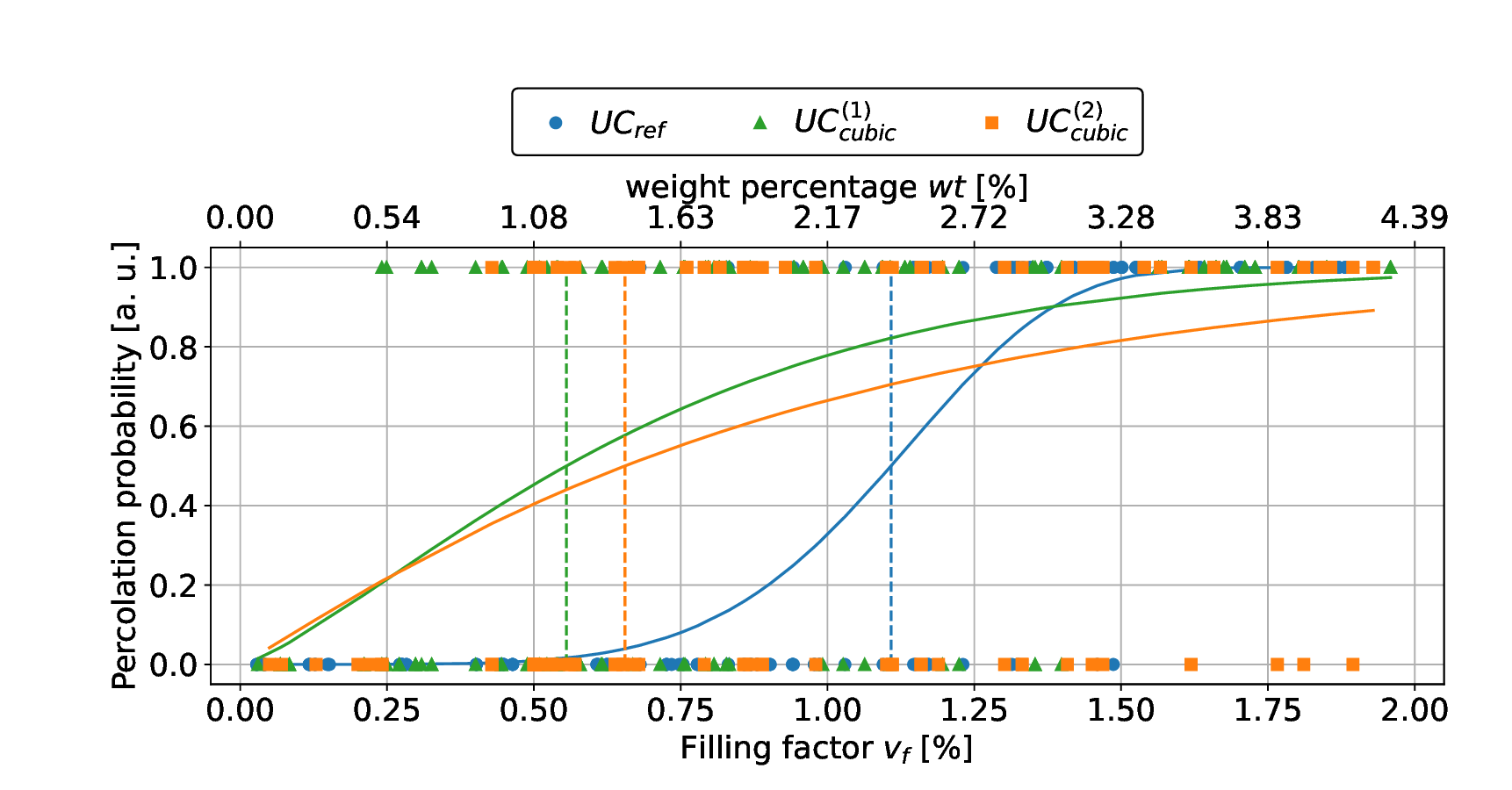}
\caption{The converted simulations results are shown in relation to $v_f$ and $wt$. Blue circles indicate data which is obtained with the reference unit cell. Converted results from $\text{UC}^{(1)}_{\text{cubic}}$ and $\text{UC}^{(2)}_{\text{cubic}}$ are plotted as green triangles and orange squares, respectively. The Weibull CDFs are shown as solid lines in the corresponding color. Dashed vertical lines indicate the percolation threshold for each unit cell.}
\label{fig_percscubic}
\end{figure}

For each unit cell size, the parameters of a Weibull CDF (Eq. \ref{eq_CDF}) are determined by a least square fit to the converted data. The resulting Weibull CDFs allow for the calculation of the percolation threshold $P_{\text{th}}$ at which the percolation probability is equal to $0.5$ and which is an established parameter in percolation theory to characterize phase transitions \cite{Simoneau2015}. For the reference unit cell, $P_{\text{th}}$ is observed at a filling factor around 1.11 \% (2.41 $wt$ $\%$) and which matches the results from other publications, such as Simoneau et al. \cite{Simoneau2015}. For the cubic unit cells $\text{UC}^{(1)}_{\text{cubic}}$ and $\text{UC}^{(2)}_{\text{cubic}}$, the percolation threshold is obtained much earlier at filling factors 0.56 \% (1.22 $wt$ $\%$) and 0.66 \% (1.43 $wt$ $\%$), respectively. The parameters of the Weibull CDFs as well as the related percolation thresholds are summarized in Tab. \ref{tab_WeibullParas} at the end of this section.

\begin{figure}[H]
\centering
\includegraphics[width=\linewidth]{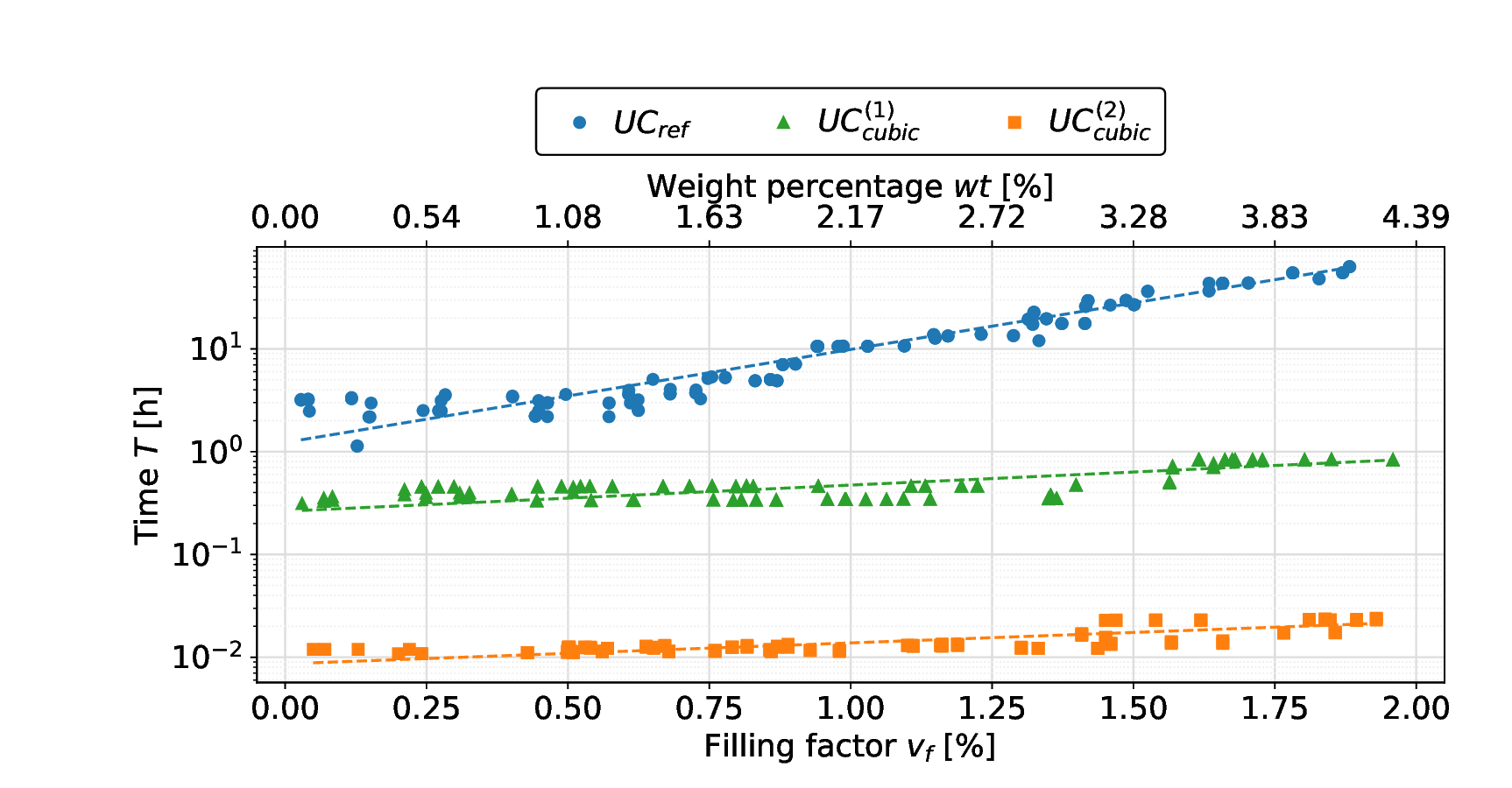}
\caption{Computation time to perform all steps with the cubic unit cells $\text{UC}_{\text{ref}}$ (blue circles), $\text{UC}^{(1)}_{\text{cubic}}$ (green triangles), and $\text{UC}^{(2)}_{\text{cubic}}$ (orange squares) as a function of the filling factor and the respective weight percentage. The fit of the exponential function (Eq. \ref{eq_timefit}) to each time data is illustrated as dashed line in the respective color.}
\label{fig_timecubic}
\end{figure}

To show that, by the use of unit cells in the model, the computation times $T$ can be reduced, they were recorded and the results are shown in Fig. \ref{fig_timecubic}.

The computation times are presented in a semi-log representation to highlight the exponential behaviour of the computation times as a function of the filling factor / weight percentage. Hence, an exponential function of the form 

\begin{equation}
    T(v_f) = a \cdot exp(v_f \cdot b) ,
    \label{eq_timefit}
\end{equation}

\noindent is used as a fit function, where $a$ (SI-unit: s) and $b$ (SI-unit: 1) are fitting parameters. The data, presented in Fig. \ref{fig_timecubic}, confirms the reduction of the simulation time by the usage of smaller unit cells. However, the obtained simulation results with $\text{UC}^{(1)}_{\text{cubic}}$ and $\text{UC}^{(2)}_{\text{cubic}}$ do not reproduce the results from $\text{UC}_{\text{ref}}$. The results suggest that the ratio of unit cell height and tube length is crucial for the percolation probability and thus, cubic unit cells are not preferable to substitute $\text{UC}_{\text{ref}}$.

\subsection{Tetragonal unit cells}
\begin{figure}[H]
\centering
\includegraphics[width=\linewidth]{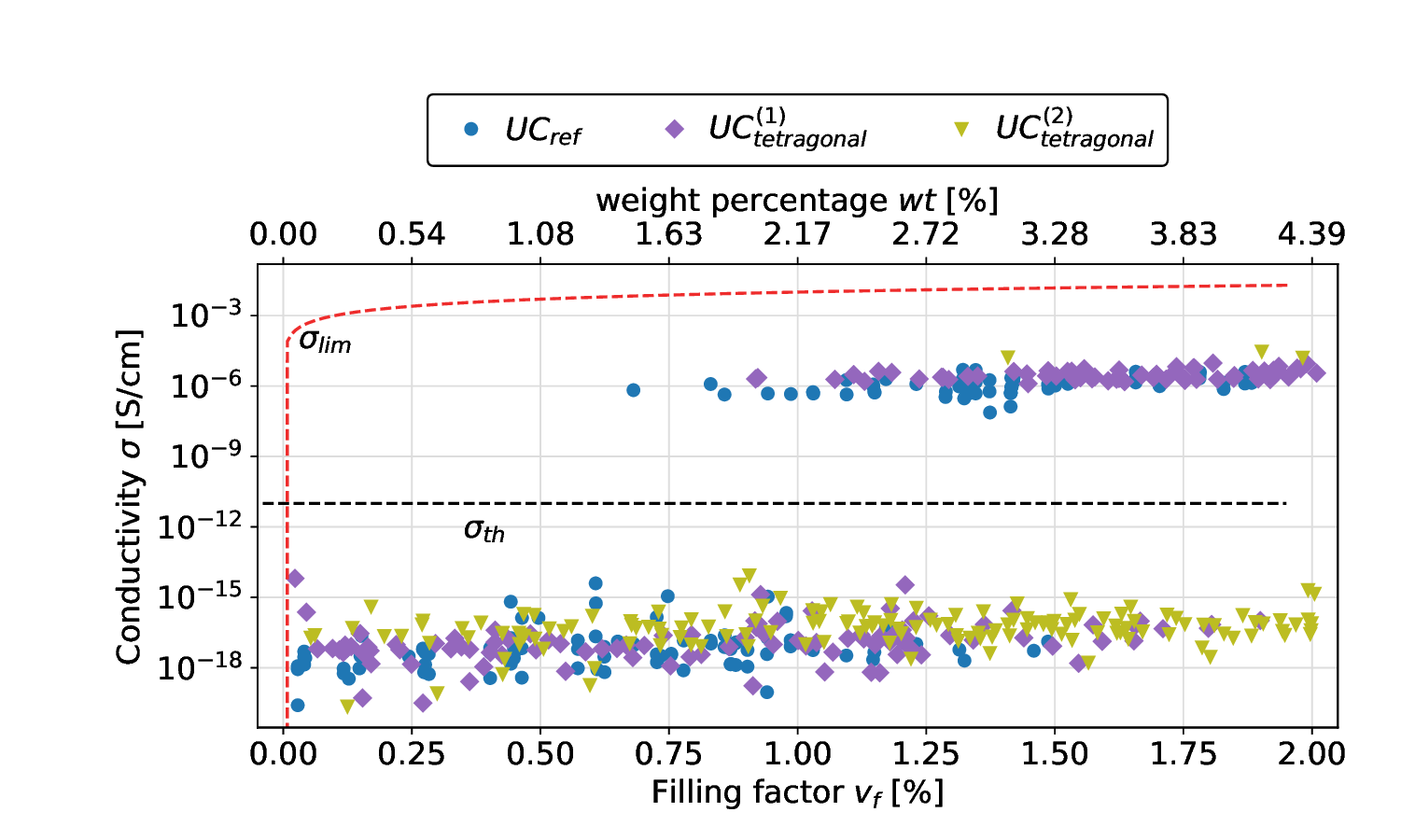}
\caption{Simulated conductivity values for the investigated tetragonal unit cell sizes $\text{UC}^{(1)}_{\text{tetragonal}}$ (purple diamonds) and $\text{UC}^{(2)}_{\text{tetragonal}}$ (yellow triangles) and for the reference unit cell $\text{UC}_{\text{ref}}$ (blue circles) as a function of $v_f$ and $wt$, respectively. The upper theoretical limit to the conductivity $\sigma_{\text{lim}}$ is illustrated by the red dashed line. A black dashed horizontal line depicts the threshold $\sigma_{\text{th}}$.}
\label{fig_condscuboid}
\end{figure}

All unit cells, studied in this section, have a height of 2500 nm. Simulation results, obtained with $\text{UC}_{\text{ref}}$, are used as reference again. The dimensions of the new tetragonal unit cells $\text{UC}^{(1)}_{\text{tetragonal}}$ and $\text{UC}^{(2)}_{\text{tetragonal}}$ are 1250 nm x 1250 nm x 2500 nm and 500 nm x 500 nm x 2500 nm, respectively. The results of the electrostatic FEM simulation is shown in Fig. \ref{fig_condscuboid}.

In accordance with the studies with the cubic unit cells, the filling factor ranges from 0.0 \% (0.0 $wt$ \%) to 2.0 \% (4.39 $wt$ \%). Even for the highest filling grades, the nanotube networks, generated with $\text{UC}^{(2)}_{\text{tetragonal}}$, do not reliably connect top and bottom electrode. The reason for this behaviour is the relative large volume of a single tube for smaller unit cell volumes (similar to the discussion of the results with $\text{UC}^{(2)}_{\text{cubic}}$) in addition with the necessity of at least one junction between two nanotubes. Simulation results, obtained with $\text{UC}^{(1)}_{\text{tetragonal}}$, indicate a transition phase in the range of 0.9 \% (1.95 $wt$ \%) to 1.7 \% (3.69 $wt$ \%). Before this phase, no interconnecting network is present. For $v_f$ values above 1.7 \%, it is very likely that the nanotube networks form a percolation path and thereby resulting in a high conductivity. Applying the threshold $\sigma_{\text{th}}$ to the data gives information about the presence or absence of a percolation path. The parameters of the Weibull CDF are determined by a least square fit to each converted data. 

\begin{figure}[H]
\centering
\includegraphics[width=\linewidth]{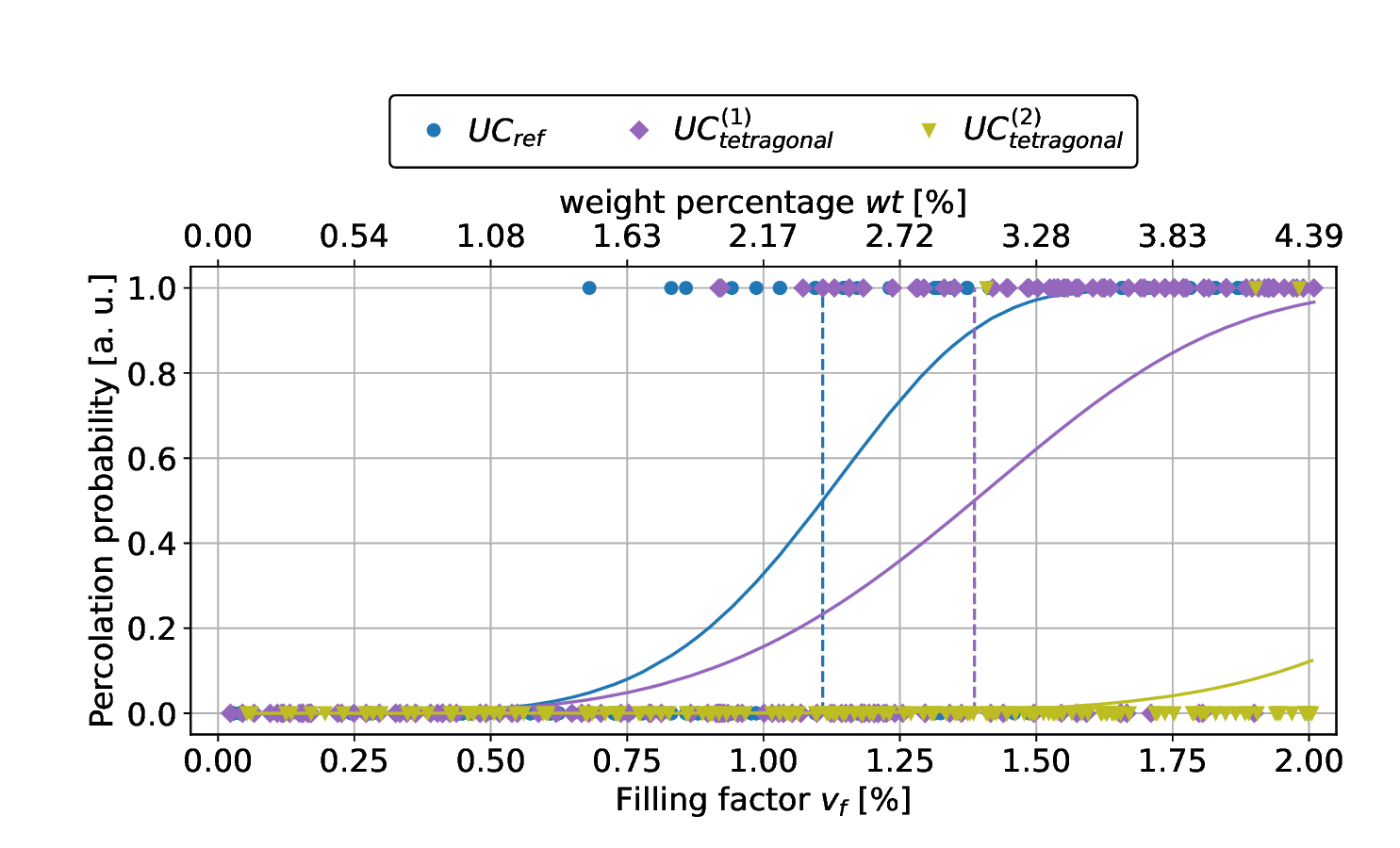}
\caption{For $\text{UC}^{(1)}_{\text{tetragonal}}$, $\text{UC}^{(2)}_{\text{tetragonal}}$, and $\text{UC}_{\text{ref}}$ the converted conductivities are shown as a function of the filling factor and the respective weight percentage. For each unit cell, the Weibull CDFs and percolation thresholds are illustrated as solid line and dashed lines in the corresponding color, respectively.}
\label{fig_perkcuboid}
\end{figure}

The percolation threshold with $\text{UC}_{\text{ref}}$ is around a filling factor of 1.11 \% (2.41 $wt$ \%). For $\text{UC}^{(1)}_{\text{tetragonal}}$ the Weibull CDF provides a similar shape and $P_{\text{th}}$ is found to be 1.39 \% (3.02 $wt$ \%). Even though the Weibull CDF of $\text{UC}^{(1)}_{\text{tetragonal}}$ and $\text{UC}_{\text{ref}}$ do not match perfectly, they are close enough for practical applications and thus can be used to substitute the reference unit cell in further studies. Within the investigated range of the filling factor, almost no percolation was observed when $\text{UC}^{(2)}_{\text{tetragonal}}$ was used. As a consequence, the parameters of the Weibull CDF as well as the percolation threshold are very different from the values obtained with $\text{UC}_{\text{ref}}$.

$\text{UC}^{(1)}_{\text{tetragonal}}$ is identified as a potential replacement for $\text{UC}_{\text{ref}}$, it raises the question of whether its use leads to a reduction in the computation time.

\begin{figure}[H]
\centering
\includegraphics[width=\linewidth]{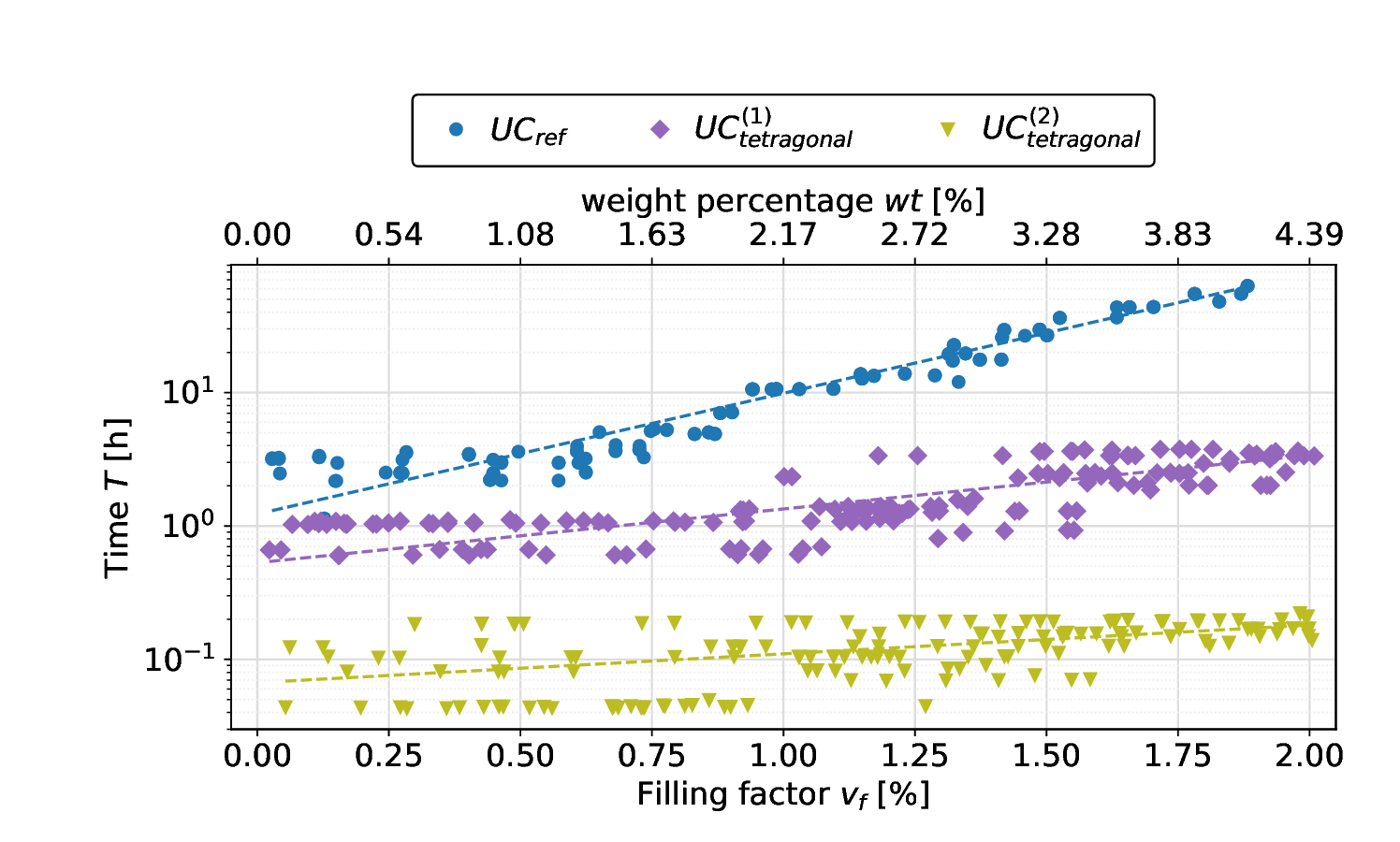}
\caption{Computation times for simulations with the unit cells $\text{UC}^{(1)}_{\text{tetragonal}}$ (purple diamonds), $\text{UC}^{(2)}_{\text{tetragonal}}$ (yellow triangles), and $\text{UC}_{\text{ref}}$ (blue circles) as a function of the filling factor $v_f$ and corresponding weight percentage $wt$. Dashed lines indicate the fit functions (Eq. \ref{eq_timefit}) to the respective time data.}
\label{fig_timecuboid}
\end{figure}

In Fig. \ref{fig_timecuboid}, the computation times for the investigated unit cells in this section are compared. For all unit cell shapes, the times show an exponential behaviour. 
While the computation times, obtained with $\text{UC}^{(1)}_{\text{tetragonal}}$ and $\text{UC}^{(2)}_{\text{tetragonal}}$, seem to exhibit a similar behaviour (except for an offset), a steeper slope for the computation time with $\text{UC}_{\text{ref}}$ is observed in Fig. \ref{fig_timecuboid}. The influence of the unit cell size and shape on the computation time is not further discussed in this work, but will be investigated in future studies.

The results, shown in Fig. \ref{fig_perkcuboid}, strongly suggest that $\text{UC}^{(1)}_{\text{tetragonal}}$ produces very similar results as the reference unit cell and thus is able to replace $\text{UC}_{\text{ref}}$. Moreover, the data, presented in Fig. \ref{fig_timecuboid}, indicate a reduction in the necessary computation time. In order to quantify the reduction of the computation time, all fit functions are normalized with respect to the exponential fit of the computation times with $\text{UC}_{\text{ref}}$. 

\begin{figure}[H]
\centering
\includegraphics[width=\linewidth]{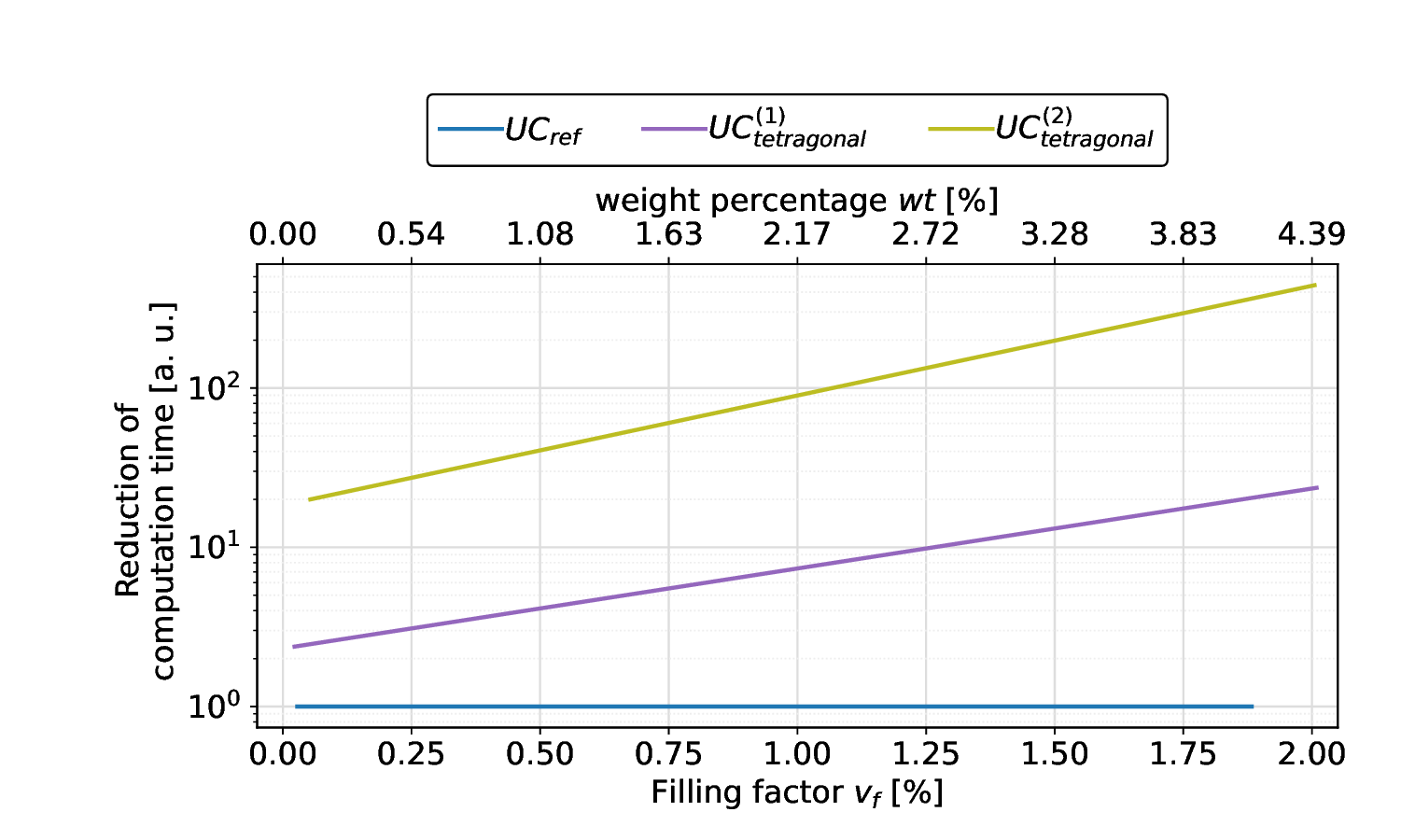}
\caption{Normalized fit functions to express the reduction in computation time. Normalizing the computation times of the reference unit cell results in a constant value, shown by the blue solid line. The time reduction for $\text{UC}^{(1)}_{\text{tetragonal}}$ and $\text{UC}^{(2)}_{\text{tetragonal}}$ is illustrated by the purple and yellow solid lines, respectively.}
\label{fig_timeenh}
\end{figure}

The normalization of the computation times show, that a significant reduction in computation time can be achieved by using $\text{UC}^{(1)}_{\text{tetragonal}}$, especially with increasing filling factor $v_f$ while receiving similar simulation results.\\
In Tab. \ref{tab_WeibullParas}, all parameters of the Weibull CDF as well as the percolation thresholds for cubic and tetragonal unit cells are summarized.

\begin{table}[H]
\centering
\caption{Weibull CDF parameters and $P_\text{th}$ for the investigated cubic and tetragonal unit cells.}
\begin{tabular}{lccccc}
Unit Cell                                   & $\lambda$       & $k$              & $P_{\text{th}}$ $(v_f)$                & $P_{\text{th}}$ $(wt)$ \\ 
\hline
 $\text{UC}_{\text{ref}}$                           & 1.19         & 5.42           & 1.11                                 &   2.41                \\
 $\text{UC}^{(1)}_{\text{cubic}}$                           & 0.73      & 1.32              & 0.56                                 &     1.22              \\
 $\text{UC}^{(2)}_{\text{cubic}}$                           & 0.92      & 1.08              & 0.66                                  & 1.43                  \\
 $\text{UC}^{(1)}_{\text{tetragonal}}$                           & 1.51        & 4.29        & 1.39                                &      3.02              \\
  $\text{UC}^{(2)}_{\text{tetragonal}}$                          & 2.55      & 8.44              & 2.44                                &   5.29                                                                  
\end{tabular}
\label{tab_WeibullParas}
\end{table}

\section{Conclusion}
The objective of the study was to examine the hypothesis that smaller unit cells have the capacity to substitute for larger unit cells. For this purpose, the percolation probability of nanotube networks as a function of the filling factor (weight percentage) with varying sizes was studied. A random-walk algorithm was developed to generate nanotubes with specified properties (e.g. length, curvature, maximal overlap when two cylinders intersect, etc.) within a predefined volume. The nanotubes display a periodic behaviour, which is of particular importance in the context of our present work. It was found that the size of the network has a significant impact on the percolation probability. The network with the largest dimensions (2500 nm x 2500 nm x 2500 nm) was regarded as the most trustful and results, which were obtained with it, were used to evaluate other network sizes. All networks were used in a numerical analysis as the computation domains. From the simulation results, the presence or absence of a percolation path in the regarding network can be derived. For each network size, the percolation probability was characterized by fitting a Weibull cumulative density function (CDF). The parameters of the Weibull CDF, along with the percolation threshold, are used to characterize the network size. The most significant factors which influence the percolation probability in the reference cell were identified and guidelines for the network size were formulated. Following these guidelines, a network size of 1250 nm x 1250 nm x 2500 nm was identified which generates results that are highly comparable to those of the reference network. This smaller network can be used as a unit cell of the reference network, since it provides a similar outcome while requiring a considerably shorter computation time. 

\section*{CRediT authorship contribution statement}
\textbf{Fabian Gumpert:} Investigation, Methodology, Visualization, Writing - Original Draft. \textbf{Dominik Eitel:} Investigation, Writing - Review $\&$ Editing. \textbf{Olaf Kottas:} Investigation, Writing - Review $\&$ Editing. \textbf{Uta Helbig:} Supervision, Resources, Writing - Review $\&$ Editing. \textbf{Jan Lohbreier:} Supervision, Resources, Writing - Review $\&$ Editing.

\section*{Declaration of competing interest}
The authors declare no competing financial interest or personal relationship that could have appeared to influence the work reported in this paper.

\section*{Data availability}
Data will be made available on request.

\section*{Acknowledgements}
The authors are grateful to the German Research Foundation (DFG) for its financial support (GZ: FIP 8/1 -2024).  

\section*{Declaration of generative AI and AI-assisted technologies in the writing process}
During the preparation of this work the authors used deepl.com in order to improve the language. After using this tool/service, the authors reviewed and edited the content as needed and take full responsibility for the content of the published article.

%Example citation, See \cite{lamport94}.

\bibliographystyle{elsarticle-num} 
\bibliography{ref}




%% For numbered reference style

%%\bibitem{lamport94}
%%  Leslie Lamport,
%%  \textit{\LaTeX: a document preparation system},
%%  Addison Wesley, Massachusetts,
%%  2nd edition,
%%  1994.

%%\end{thebibliography}
\end{document}